\documentclass[submission, PhysicsCodebases]{SciPost}

\usepackage{lineno}

\usepackage{amsmath}
\usepackage{amssymb}
\usepackage{graphicx}

\usepackage{listings}

\usepackage[normalem]{ulem} 

\usepackage{listings}
\makeatletter

\usepackage{jlcode}

\newcommand{\codeurl}{https://raw.githubusercontent.com/m3g/jlcode_example/master}
\newcommand{\codelinktext}{[Click here to download the code]}

\def\jlbasicfont{\ttfamily\normalsize\selectfont}

\lstdefinestyle{code}{
  language=Julia,
  showstringspaces=false,
  keywordstyle={[1]\color[RGB]{255,134,2}},
  keywordstyle={[2]\color{jlliteral}},
  keywordstyle={[3]\color[RGB]{0,143,0}},
  keywordstyle={[4]\color{violet}},
  keywordstyle={[5]\color[RGB]{0,84, 232}},
  commentstyle={\color{jlcomment}},
  stringstyle={\color[RGB]{216,72,0}},
  identifierstyle={\color{jlbase}},
  columns=fullflexible,
  keepspaces=true
}

\lstdefinestyle{codenohighlight}{
  language=Julia,
  showstringspaces=false,
  keywordstyle={[1]\color{black}},
  keywordstyle={[2]\color{black}},
  keywordstyle={[3]\color{black}},
  keywordstyle={[4]\color{black}},
  keywordstyle={[5]\color{black}},
  commentstyle={\color{black}},
  stringstyle={\color{black}},
  identifierstyle={\color{black}},
  columns=fullflexible,
  keepspaces=true
}

\lstnewenvironment{jlcodeblock}[1][]{
  \lstset{style=code,xleftmargin=0.2cm,xrightmargin=0.2cm,#1}
}{}
\lstnewenvironment{jlcodeblocknohighlight}[1][]{
  \lstset{style=codenohighlight,xleftmargin=0.2cm,xrightmargin=0.2cm,#1}
}{}

\newcommand{\codelink}[1]{\lstinputlisting[style=code]{#1}
  \noindent\begin{center}
  \filename@parse{#1}
  \href{\codeurl/\filename@base.\filename@ext}
  {\textcolor{blue}{\codelinktext}}
  \end{center}
}
\newcommand{\linkonly}[1]{
  \protect\filename@parse{#1}
  \href{\codeurl/\filename@base.\filename@ext}
  {\textcolor{blue}{\tt[\protect\filename@base.\protect\filename@ext]}}
}

\newcommand{\jlinlcustom}[1]{%
\settowidth{\jlinlem}{\jlinlfont{m}}%
\lstinline[language=julia,style=code,
basicstyle=\jlinlfont,basewidth=\jlinlem]^^a7#1^^a7}

\newcommand{\inline}{\jlinlcustom}

\newcommand{\jlinlcustomnohighlight}[1]{%
\settowidth{\jlinlem}{\jlinlfont{m}}%
\lstinline[language=julia,style=codenohighlight,
basicstyle=\jlinlfont,basewidth=\jlinlem]^^a7#1^^a7}

\newcommand{\inlinenohighlight}{\jlinlcustomnohighlight}

\newcommand{\code}[1]{\verb|\mbox{#1}|}
\newcommand{\ket}[1]{|#1\rangle}
\newcommand{\bra}[1]{\langle#1|}

\begin{document}

\begin{center}{\Large \textbf{
The ITensor Software Library for Tensor Network Calculations
}}\end{center}

\begin{center}
Matthew Fishman\textsuperscript{1},
Steven R. White\textsuperscript{2},
E. Miles Stoudenmire\textsuperscript{1*}
\end{center}

\begin{center}
{\bf 1} Center for Computational Quantum Physics, Flatiron Institute, New York, NY 10010, USA
\\
{\bf 2} Department of Physics and Astronomy, University of California, Irvine, CA 92697-4575 USA
\\
* mstoudenmire@flatironinstitute.org
\end{center}

\begin{center}
\today
\end{center}


\section*{Abstract}
{\bf
ITensor is a system for programming tensor network calculations with an interface modeled on 
tensor diagrams, allowing users to focus on the connectivity of a tensor network without manually bookkeeping tensor indices. The ITensor interface rules out common programming errors and enables rapid prototyping of algorithms. After discussing the philosophy behind the ITensor approach, we show examples of each part of the interface including Index objects, the ITensor product operator, tensor factorizations, tensor storage types, algorithms for matrix product state (MPS) and matrix product operator (MPO) tensor networks,  quantum number conserving block sparse tensors, and the NDTensors library. We also review publications that have used ITensor for quantum many-body physics and for other areas where tensor networks are increasingly applied. To conclude we discuss promising features and optimizations to be added in the future.
}

\vspace{10pt}
\noindent\rule{\textwidth}{1pt}
\tableofcontents\thispagestyle{fancy}
\noindent\rule{\textwidth}{1pt}
\vspace{10pt}

\section{Introduction}

Tensor networks are a technique for working with tensors which have many 
indices \cite{Orus:2014a,Cichocki:2016,Schollwoeck:2011,Ostlund:1995,Hackbusch:2009,Oseledets:2011}.
The naive memory and computing costs of working with a tensor having $N$ indices (an order-$N$ tensor) 
scales exponentially with $N$.
A \emph{tensor network} is a representation of a large, high-order tensor as the contracted
product of many low-order tensors.
When all of the tensors in the network are low-order, a tensor network can make it 
efficient to perform important operations such as summing two high-order tensors or
computing their inner product. These operations can remain efficient whether the 
high-order tensor represented implicitly by the network has hundreds, thousands, or even an infinite number of indices. 

Describing tensor networks can be difficult when using traditional notation: one must come up with distinct names for indices and it can be hard to see the  connectivity pattern of the network. An elegant alternative is tensor diagram notation \cite{Penrose}. In diagram notation, tensors are shapes and indices are depicted as lines emanating from them. Connecting two index lines means they
are contracted or summed over. For example, the following diagram is equivalent to the traditional expression below it:
\begin{center}
\includegraphics[width=0.5\columnwidth]{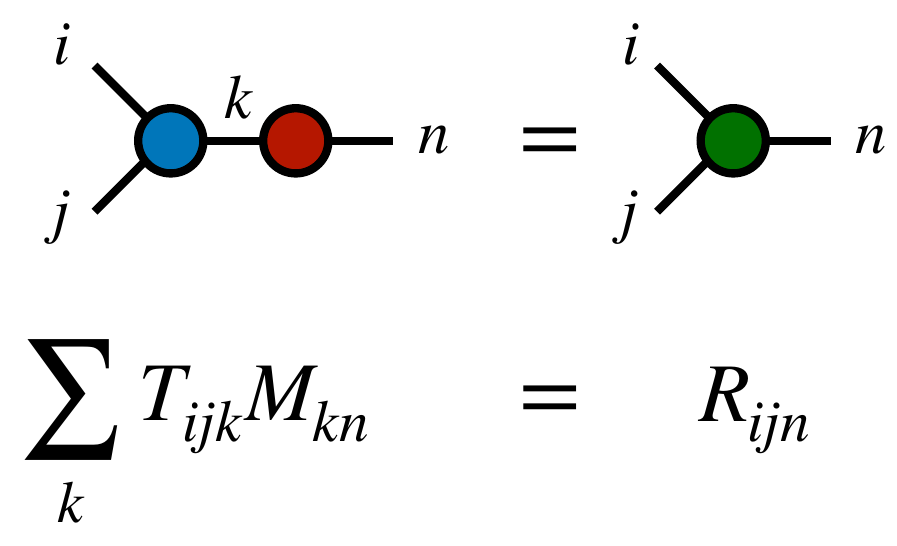}
\end{center}

Diagram notation is enormously helpful for expressing tensor networks, as it emphasizes key aspects of tensor algorithms while suppressing implementation details such as the ordering of tensor indices. 
It is just as rigorous as traditional index notation.

ITensor, short for \emph{intelligent tensor}, is a software library inspired by tensor diagram notation. Its goal is enabling users to translate a tensor diagram into code without reintroducing concepts not expressed by tensor diagrams.
For example, when summing two ITensors the only requirement is that they have the same set of indices in any order; the ITensor system handles all other details of performing the sum.

Two ``philosophical'' principles guided the design of ITensor.
The first was that in using the library, any implementation details which are not a part of the conceptual algorithms should be kept hidden from the user as much as possible. Not having to think about these details allows one to focus more clearly on the essentials. A key early insight was that this principle could apply to the ordering of indices in an ITensor.  In typical tensor software, the user is constantly thinking about the order of the indices. However, tensor  diagrams do not have any index ordering, just labels that keep track of the relevant information. Thus ITensors have the ordering of their indices abstracted away as an implementation detail, using ``intelligent'' indices that retain their identity.  The second key principle is that the software should allow one to interact with it at a variety of levels. At a high level, for calculations done in a standard way one can call functions encapsulating a sophisticated algorithm (say, the density matrix renormalization group, DMRG) without understanding much of the implementation details. At an intermediate level, one can gain flexibility by working with moderately sophisticated routines, such as for adding MPS. And finally, to do something more novel, one can work at the lower level of individual ITensors. This multilevel access mandated that ITensor be a library, not a single executable program with complicated input files.

These initial principles led to other interesting design choices over time. One consequence of having intelligent tensor indices as distinct data objects (of type \inlinenohighlight{Index}) is that they can store extra information about themselves. A key use case is indices which have internal subspaces labeled by conserved quantum numbers (symmetry group representation labels). Storing this information in the \inlinenohighlight{Index} objects ensures that when contracting two quantum number conserving ITensors, both are guaranteed to use the same ordering of the subspaces for storing their data. Another consequence is that dense and sparse ITensors can be of the same type and have essentially the same interface, because the implementation can inspect the indices and internal storage type to determine the actual `type' of any ITensor. Thus users can write very generic code that works for any type of ITensor.

The design choice to have an ITensor manage its own index ordering is by no means an obvious one. Benefits of ITensor's \emph{intelligent index} system include making addition of ITensors $A$ and $B$ as simple as writing the code $A+B$ for tensors with the same indices, or automating the application of operators to matrix product states (MPS). A system to automate anti-commuting ``fermionic'' tensor algebras is currently in development which heavily relies on the intelligent index system to keep track of index ordering.
Calculations where multiple tensor diagrams have many of the same indices is another example where intelligent indices makes code simpler and less error prone. An example is taking the gradient of a tensor network, which simply involves removing that tensor from the network.
But possible drawbacks of the intelligent index approach include occasional extra lines of code to manipulate index properties and some loss of control over low-level details of tensor operations when using very high-level features. 
However we do offer more advanced features that give complete control over such details.

Most tensor libraries, in contrast, choose to expose the ordering of tensor indices to users who must manage this ordering manually \cite{Psarras:2021}.  Such interfaces always give users fine-grained control over details that can affect performance, but can put more of a burden on the user to ensure correctness. While ITensor does not require users to think about the index ordering, it can be manually controlled when needed by calling functions such as \inlinenohighlight{permute} to explicitly permute indices into a specified memory ordering. In tensor contractions, the index ordering of the output tensor can be controlled by supplying it through an in-place contraction function. 

Another contrast between ITensor and other tensor libraries relates
to how networks consisting of many tensors are handled. In many
libraries, higher-level network interfaces are offered by supplying
temporary text labels for indices
\cite{Pfeifer:2015,Haegeman:2021,Harris:2020}
or by placing tensors into a graph or network structure and specifying
contracted indices through the graph topology \cite{Gray:2018,Roberts:2019}.
Because ITensors have persistently labeled indices, any collection
of ITensors with unique indices already specifies a graph.
We are currently taking advantage of this property of ITensors to
offer higher-level tensor network abstractions and make use of it
in our upcoming automatic differentiation tools.

ITensor was first implemented in C++ and extensively developed and refined
through three major releases over ten years.
\footnote{\mbox{ITensor Github Repository (C++):} 
\href{https://github.com/ITensor/ITensor}{https://github.com/ITensor/ITensor}} 
Recently, ITensor has been fully ported to the Julia language and most new
features are being developed there.
\footnote{\mbox{ITensor Github Repository (Julia):} 
\href{https://github.com/ITensor/ITensors.jl}{https://github.com/ITensor/ITensors.jl}} 
In what follows we show examples in Julia, though we emphasize that the 
high-level C++ and Julia interfaces are quite similar (see the Appendix for full 
code examples in each language). 
Both versions are full implementations of ITensor in each language: 
the Julia version is not a wrapper around the C++ version.

The goal of this article is to provide a high-level overview of the ITensor system, its design goals,
and its main features. 
Much more information including detailed documentation of the ITensor interface, code examples,
and tutorials can be found on the ITensor website: {\color{blue} \href{https://itensor.org/}{https://itensor.org}}.

\section{Interface Examples}

We first introduce ITensor by giving examples as an informal overview.
In later sections, we will discuss many more details of the individual elements
making up the ITensor system such as ``intelligent'' tensor indices, tensor 
factorizations, and block sparse ITensors.

\subsection{Installing ITensor}

Julia features a built-in package manager that makes installing libraries simple.
To install the ITensor library, all a user has to do is issue the following commands,
starting from their terminal:
\begin{jlcodeblock}
$ julia
julia> ]
pkg> add ITensors
\end{jlcodeblock}
The \inlinenohighlight{julia} command starts an interactive Julia session and typing \inlinenohighlight{]} enters package
manager mode. The command \inline{add ITensors} downloads and installs all the 
dependencies of the \href{https://github.com/ITensor/ITensors.jl}{ITensors.jl} package
then finally the ITensor library itself.\footnote{The reason the Julia library is called ``ITensors''
and not ITensor is to keep the module name from conflicting with name of the ITensor type.}

\subsection{Obtaining Help}

Once ITensor is installed, the built-in Julia documentation system can be used to query ITensor
functions and types. For example
\begin{jlcodeblock}
julia> using ITensors
julia> ?
help?> Index
\end{jlcodeblock}
will give the output
\begin{jlcodeblocknohighlight}
search: Index indexin IndexStyle IndexLinear ...

  An Index represents a single tensor index with fixed
  dimension dim. Copies of an Index compare equal unless
  their tags are different.
  
  ...
\end{jlcodeblocknohighlight}
and additional information describing the \inlinenohighlight{Index} type and its constructors.

\subsection{Basic ITensor Usage}

To begin using the ITensors package in a Julia session or script, input
the line 
\begin{jlcodeblock}
using ITensors
\end{jlcodeblock}
Before creating an ITensor, one first creates its indices. 
The line of code
\begin{jlcodeblock}
i = Index(3)
\end{jlcodeblock}
creates a tensor Index of dimension 3 and assigns this Index object to the reference
\inlinenohighlight{i}.
Upon creation, this Index is stamped with an immutable, unique id number which allows copies of
the Index to be compared and matched to one another. 
A portion of this id is shown when printing the Index, with typical example output 
of the command \inlinenohighlight{@show i} being:
\begin{jlcodeblock}
i = (dim=3|id=804)
\end{jlcodeblock}

After making a few Index objects \inlinenohighlight{i,j,k,l} one can define ITensors:
\begin{jlcodeblock}
A = ITensor(i)
B = ITensor(j,i)
C = ITensor(l,j,k)
\end{jlcodeblock}
Because matching Index pairs can automatically recognize each other through their id numbers, 
tensor contraction
can be carried out as:
\begin{jlcodeblock}
D = A * B * C
\end{jlcodeblock}
The \inlinenohighlight{*} operator finds all matching indices between two ITensors and 
sums over or contracts these indices. The \inlinenohighlight{i} Index is summed in the first
contraction above and \inlinenohighlight{j} in the second, leaving \inlinenohighlight{D} with indices \inlinenohighlight{l} and \inlinenohighlight{k}.
The ITensor product operator ``\inlinenohighlight{*}'' can also be used for outer products and scalar products,
and is discussed in more detail in Section~\ref{sec:product}.

\subsection{Setting ITensor Elements}
Setting an element of an ITensor \inline{A = ITensor(i,j,k)} is done by 
\begin{jlcodeblock}
A[i=>2,j=>3,k=>1] = 0.837
\end{jlcodeblock}
which assigns the value 0.837 to the element of \inlinenohighlight{A} for which index \inlinenohighlight{i} 
has the value 2, \inlinenohighlight{j} has value 3, and \inlinenohighlight{k} has value 1.
(Note that in Julia, the built-in notation \inline{x=>y} makes a \inline{Pair(x,y)}
object.) ITensor indices are 1-indexed, similar to Julia arrays.

Because the Index objects are provided along with their values,
they can be passed in any order. Thus the following lines of code
\begin{jlcodeblock}
A[i=>2,j=>3,k=>1] = 0.837
A[k=>1,i=>2,j=>3] = 0.837
\end{jlcodeblock}
have exactly the same effect on the ITensor \inlinenohighlight{A}.

To create an ITensor with normally-distributed random elements instead
of specific values, one can use the constructor
\begin{jlcodeblock}
T = randomITensor(i,j,k)
\end{jlcodeblock}
to make a real-valued random tensor or
\begin{jlcodeblock}
T = randomITensor(ComplexF64,i,j,k)
\end{jlcodeblock}
to construct a complex-valued random ITensor.

\subsection{Matrix Example}
To illustrate the usefulness of the ITensor approach involving Index objects
and the \inlinenohighlight{*} operator, 
consider a pair of order-2 tensors (matrices)
\begin{jlcodeblock}
A = ITensor(i,j)
B = ITensor(k,j)
\end{jlcodeblock}
In a typical matrix or tensor library, to contract \inlinenohighlight{A} with \inlinenohighlight{B} and sum
over their shared index \inlinenohighlight{j}, one would need to write code similar to
\begin{jlcodeblock}
C = A * transpose(B)
\end{jlcodeblock}
Note that the above line is \emph{not} ITensor code!

Within ITensor, all one needs to do is to write
\begin{jlcodeblock}
C = A * B
\end{jlcodeblock}
and the \inlinenohighlight{*} operator handles the transposition of \inlinenohighlight{B} automatically.
If \inlinenohighlight{B} is redefined with the ordering of its indices reversed, the operation
\inlinenohighlight{A * B} continues to give the correct result. This type of behavior makes ITensor 
applications robust to changes in the code that may modify the ordering of tensor indices
or the layout of tensors in memory.

\subsection{Summing ITensors}

ITensors can be added and subtracted as long as they have the same set of indices. Even if the indices are in a  different order, addition always works straightforwardly because the ITensor system is able to internally deduce the data permutation required:

\begin{jlcodeblock}
A = randomITensor(i,j,k)
B = randomITensor(k,i,j)
C = A + B	
\end{jlcodeblock}

ITensors may also be subtracted and multiplied by scalars, including complex scalars,
for example:
\begin{jlcodeblock}
D = 4*A - B/2
F = A + 3.0im * B
\end{jlcodeblock}

\subsection{Priming Indices}

Sometimes it is not desirable to contract all of the indices shared between two tensors. Consider two ITensors
\begin{jlcodeblock}
A = ITensor(i,j)
B = ITensor(i,j)	
\end{jlcodeblock}
and say we want to contract only over the index \inlinenohighlight{j} leaving the \inlinenohighlight{i} indices uncontracted.

A convenient way to achieve this while still using the \inlinenohighlight{*} operator is to \emph{prime} one of the \inlinenohighlight{i} indices
\begin{jlcodeblock}
Ap = prime(A,i)
\end{jlcodeblock}
The ITensor \inlinenohighlight{Ap} has the same elements as \inlinenohighlight{A} 
but has indices \inline{(i',j)}. When contracting \inlinenohighlight{Ap}
with \inlinenohighlight{B}, now only the \inlinenohighlight{j} indices will match or compare equal, so it will
be the only Index contracted
\begin{jlcodeblock}
C = Ap * B
hasind(C,i) == true
hasind(C,i') == true
\end{jlcodeblock}
Diagrammatically we can notate the above contraction as:
\begin{center}
\includegraphics[width=0.5\columnwidth]{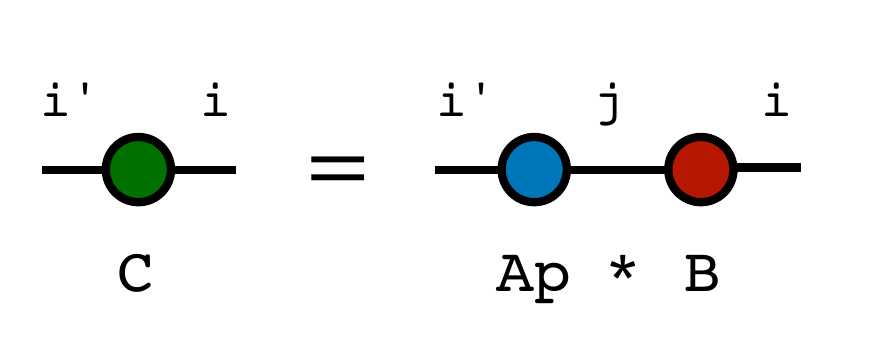}
\end{center}

\subsection{Compiling ITensor}

Although the experience of using Julia is similar to using an interpreted language,
it is actually a \emph{just-in-time compiled} language. 

The initial compilation time when Julia first encounters new functions or types can be large in a new Julia session, though 
there is ongoing work to provide ahead-of-time compilation tools for Julia.
To reduce just-in-time compilation overhead, we offer a convenient way for users to compile
most of the ITensors.jl code ahead of time with the following commands within an 
interactive Julia session:
\begin{jlcodeblock}
julia> using ITensors
julia> ITensors.compile()
\end{jlcodeblock}
The compilation process can take many minutes, but only has to be performed once each
time the ITensors.jl library is upgraded to a new version. 
After the command is run, it will suggest command-line arguments
that can be passed to the \inlinenohighlight{julia} language program
that will load a precompiled ITensors.jl system image when running Julia. Running ITensor
code this way typically reduces startup times to only a few seconds.

\subsection{Online Code Examples}

For more extensive and frequently updated examples of ITensor code, including full
applications, we include an set of examples as part of our source code distribution at the 
following link: {\color{blue} \href{https://github.com/ITensor/ITensors.jl/tree/main/examples}{ITensor Code Examples}}.

\section{Index Objects}

A core concept of the ITensor system is that tensor indices 
carry information beyond just their dimension. Mathematically, 
this corresponds to the notion that an index labels the 
basis of a vector space, and that two vector spaces may be 
different from each other despite having the same dimension.

The notion that a tensor index corresponds to a specific
vector space is encoded in the unique \emph{id number} assigned to an 
Index object when it is constructed:
\begin{jlcodeblock}
i = Index(4)
@show i  # prints: i = (dim=4|id=577)
\end{jlcodeblock}
Printing an Index as in the code above shows a portion of 
the (64 bit) id number.\footnote{As a technical note, the Index id numbers are 
generated randomly, but collisions are highly improbable because of 
the 64-bit length of the ids. Random id generation has many advantages
over sequential, including using ITensor for parallel algorithms and 
reading ITensors from files and mixing these ITensors with newly
generated ones.}

Because a new id is assigned each time an Index is constructed, 
other separately constructed Index objects 
will not compare equal to \inlinenohighlight{i} even if they have the same dimension
\begin{jlcodeblock}
j = Index(4)
j != i	# true
\end{jlcodeblock}
In other words, comparison operations (\inlinenohighlight{==},\inlinenohighlight{!=}) require two Index objects to have the same id for them to compare equal.

To enrich the Index system one may also add tags to indices
\begin{jlcodeblock}
s = Index(3,"s,Site")	
\end{jlcodeblock}
The Index \inlinenohighlight{s} above has a dimension 3, as well as two 
tags \inlinenohighlight{"s"} and \inlinenohighlight{"Site"}.
For efficiency reasons, tags can have a maximum of eight characters and indices can have a maximum of four tags.
These maximum values are currently hard-coded into the library and may be increased in the future as use cases arise that require longer tags or more tags.

Tags can serve multiple purposes: helping to identify Index 
objects when printing them; 
collecting subsets of indices sharing a common tag or tags;
and preventing certain Index pairs from contracting with each other. 
This last use of tags extends the rule for
Index comparisons: for Index objects to compare equal they
must have the same tags as well as the same id number.

As discussed in the previous section, one other way to prevent 
Index objects from comparing equal is to
change their \emph{prime level}. Every Index carries an integer
prime level which defaults to zero. 
\begin{jlcodeblock}
i = Index(2,"i")
@show plev(i) # plev(i) = 0
\end{jlcodeblock}

A copy of Index \inlinenohighlight{i} but with 
a prime level of~1 can be created by calling
\begin{jlcodeblock}
ip = prime(i)
@show plev(ip) # plev(ip) = 1	
\end{jlcodeblock}
or for convenience by writing
\begin{jlcodeblock}
ip = i'
\end{jlcodeblock}

Two copies of the same Index which have 
different prime levels do not compare equal
\begin{jlcodeblock}
i == i'	 # false
i == i'' # false
\end{jlcodeblock}

Because both primes and tags can be used to
prevent Index objects from comparing equal
to each other and being contracted by the
\inlinenohighlight{*} operator, some experience is needed
to choose the best approach. Primes are 
useful when indices are only distinguished 
temporarily; it is easy afterward to call
\inline{noprime(T)} on an ITensor to reset
the prime levels of all of its indices.
On the other hand, tags should be used when
there is some application-specific understanding
of why certain indices are distinguished.
For example in the case of a tensor network with a square 
lattice structure, where all indices linking the 
tensors together may describe the same vector space, 
we might use the tags  \inlinenohighlight{"x=-1"}, \inlinenohighlight{"x=0"}, \inlinenohighlight{"x=1"}, $\dots$ to label a unique horizontal position in the lattice
and the tags \inlinenohighlight{"y=-1"}, \inlinenohighlight{"y=0"}, \inlinenohighlight{"y=1"}, $\dots$ to specify a unique vertical position.
This is particularly useful in applications involving translational
invariance, where many copies of the same Index can appear
in different contexts and it can become cumbersome to distinguish
them by prime levels alone.

\section{The ITensor Product Operator (\texorpdfstring{$*$}{contract}) \label{sec:product}}

Just as tensor diagrams unify many concepts, the ITensor product operator \inlinenohighlight{*} 
likewise unifies many operations into a single operation:

\begin{itemize}
\item The \inlinenohighlight{*} product of ITensors with no indices in common computes an
\emph{outer product}.
\item The \inlinenohighlight{*} product of ITensors with all the same indices computes an
\emph{inner product}, resulting in a scalar ITensor.
\item Otherwise, for a pair of ITensors having just some indices in common, the \inlinenohighlight{*} operator
computes a \emph{tensor contraction}.
\end{itemize}

A simple example of an outer product is the product of two vectors which do not share a common index:
\begin{jlcodeblock}
v = ITensor(i)
w = ITensor(j)
x = v * w
\end{jlcodeblock}
\begin{center}
\includegraphics[width=0.5\columnwidth]{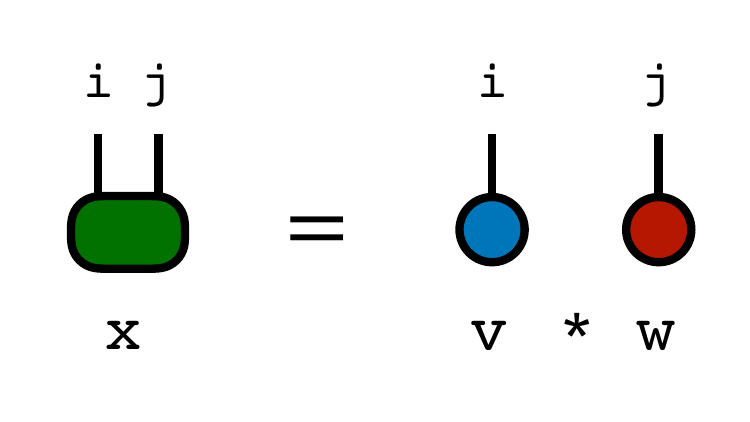}
\end{center}

Using the \inlinenohighlight{*} operator to compute an inner product results in a scalar ITensor
with no indices as in the following example (note that the indices do not need to be in the same order for the result to be correct):
\begin{jlcodeblock}
A = ITensor(i,j,k)
B = ITensor(k,i,j)
C = A * B
\end{jlcodeblock}
\begin{center}
\includegraphics[width=0.5\columnwidth]{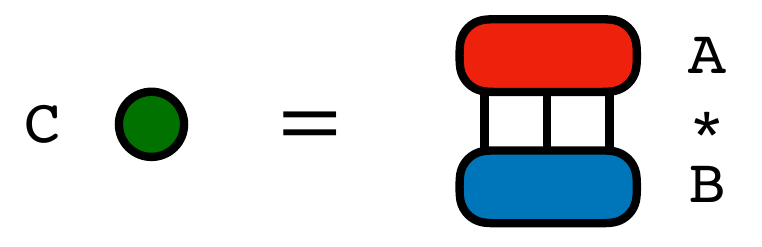}
\end{center}
The \inlinenohighlight{scalar} function can be called to convert 
a scalar ITensor into a real or complex number
\begin{jlcodeblock}
x = scalar(C)
\end{jlcodeblock}
or alternatively one can call \inline{x = C[]}.

Finally, to illustrate the case of a tensor contraction where
only some of the indices are summed, we can use the following example
which was also shown at the beginning of this article:
\begin{jlcodeblock}
T = ITensor(i,j,k)
M = ITensor(k,n)
R = T * M
\end{jlcodeblock}
\begin{center}
\includegraphics[width=0.5\columnwidth]{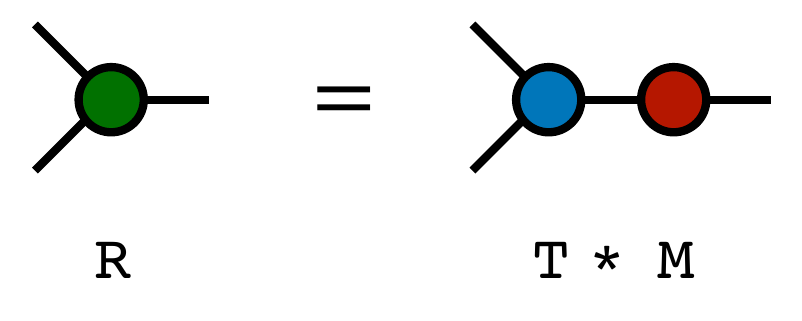}
\end{center}
In the diagram above, we have omitted the names of the indices to emphasize
the typical user experience: all that a user needs to know to get a correct
result in the above example is that \inlinenohighlight{T} and \inlinenohighlight{M} share one Index.
Keeping track of the ordering of the uncontracted indices, which become the
indices of \inlinenohighlight{R}, is not necessary.

Besides contracting regular tensors, the \inlinenohighlight{*} operator can also be used 
in conjunction with specially constructed tensors to manipulate tensor indices.
One example of such a special tensor type is a \emph{delta tensor}, 
also known as a copy tensor, which has all diagonal
elements equal to one and other elements equal to zero, and is 
often shown as a solid black circle in tensor diagrams.
In the ITensor library, a delta tensor uses special diagonal-sparse storage 
internally, not only to save memory but also to ensure that the contraction of delta
tensors with other tensors is performed using specially optimized routines.

A delta tensor can be used to replace
an Index with another Index of the same dimension:
\begin{jlcodeblock}
A = ITensor(k,j)
A = A * delta(k,i)
@show hasind(A,i) # true
\end{jlcodeblock}
\begin{center}
\includegraphics[width=0.5\columnwidth]{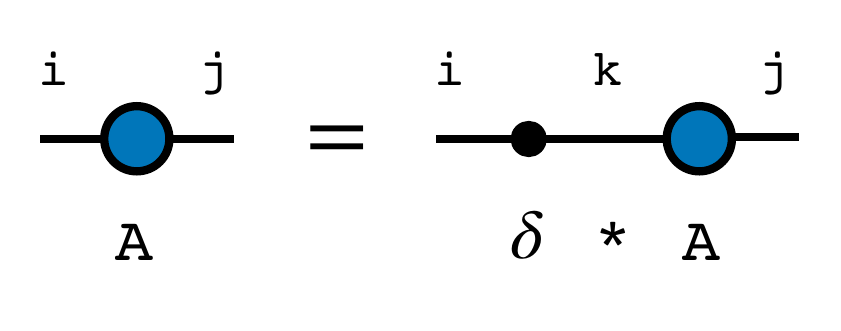}
\end{center}
or to duplicate (or split) an index as follows:
\begin{jlcodeblock}
B = ITensor(k)
B = B * delta(k,i,j)
\end{jlcodeblock}
\begin{center}
\includegraphics[width=0.5\columnwidth]{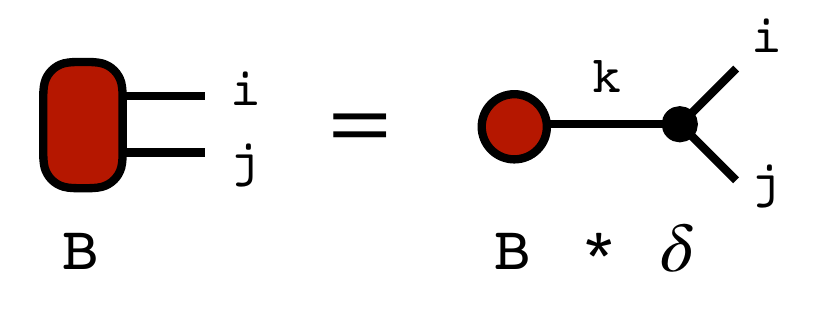}
\end{center}
Note that in Julia, one can use the unicode character
$\delta$ to write the code above as \inline{B = B * δ(k,i,j)}.

Another example of a special tensor type is a \emph{combiner} ITensor.
When contracted with another ITensor, a combiner merges multiple
indices into a single Index.
\begin{jlcodeblock}
T = ITensor(i,j,k)
C = combiner(i,j)
cT = C * T
\end{jlcodeblock}
\begin{center}
\includegraphics[width=0.5\columnwidth]{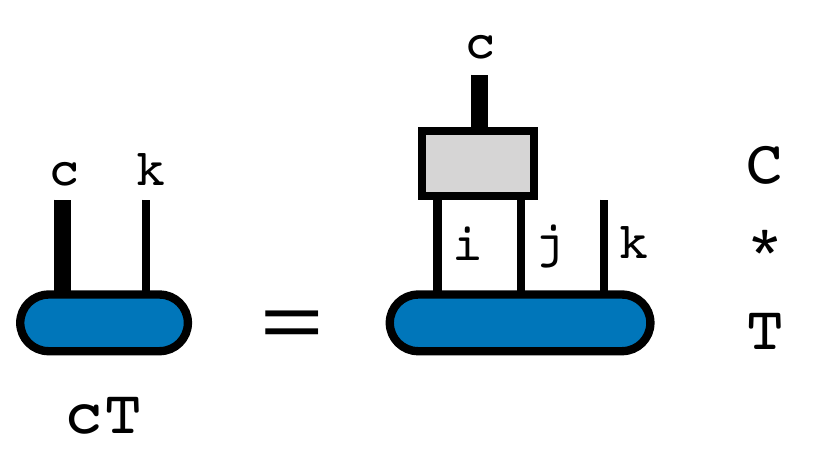}
\end{center}
The Index \inlinenohighlight{c} shown in the diagram above can be retrieved by
calling \inline{combinedind(C)} on the combiner ITensor. 
Alternatively one can call \inline{commonind(C,cT)} to retrieve this Index, since it is the one that
the combiner and \inlinenohighlight{cT} will necessarily have in common.

Taking the product with the conjugate of the combiner reverses this operation.
\begin{center}
\includegraphics[width=0.6\columnwidth]{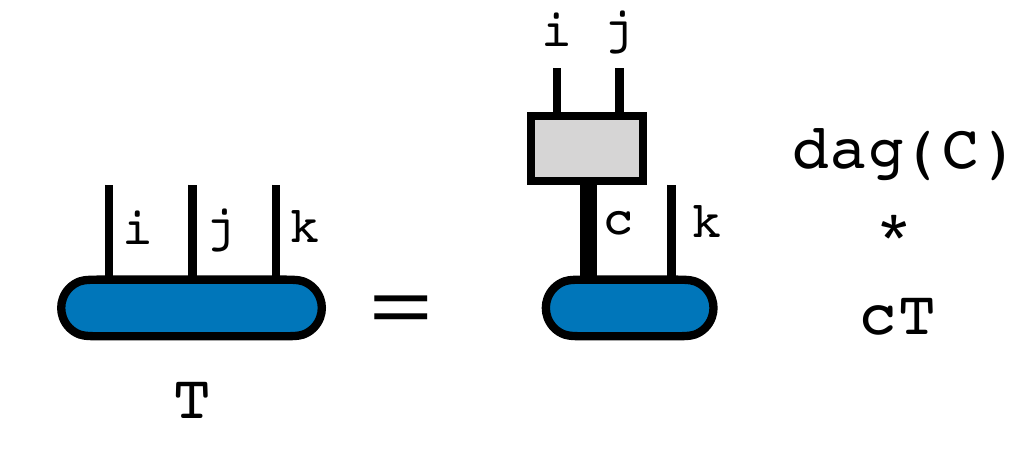}
\end{center}
Like delta tensors, combiners also use a special storage type with 
a negligible memory footprint and optimized contraction algorithms for combining and uncombining indices.

The action of a combiner on a tensor is conceptually identical to the concept of permuting and reshaping a 
multi-dimensional array, at least for the case of dense ITensors. For quantum number conserving or symmetric
ITensors, combiners can perform additional steps like grouping multiple copies of a quantum number together
in the combined Index, or managing anticommutation properties in the case of the upcoming ITensor fermion system.

\section{Tensor Decompositions}

Many commonly used tensor network decompositions are built from 
matrix decompositions such as the QR and singular value decompositions (SVD) known from linear algebra.
Despite being defined in terms of matrices, these factorizations 
can be straightforwardly defined for tensors too. All that is 
needed is a mapping from a tensor to a matrix, defined by 
specifying a certain group of indices as row indices and the rest as column 
indices, then treating each group as a single larger index when computing 
the decomposition.
ITensor automates the tedious and error-prone process of converting
tensors to matrices and back, providing a tensor-level interface for
various decompositions.

Consider an ITensor \inlinenohighlight{T} with indices \inlinenohighlight{i,j,k}. We can
compute a QR decomposition of \inlinenohighlight{T} by just specifying that
\inlinenohighlight{i,k} are the row indices as follows:
\begin{jlcodeblock}
T = randomITensor(i,j,k)
Q,R = qr(T,(i,k))
\end{jlcodeblock}
\begin{center}
\includegraphics[width=0.6\columnwidth]{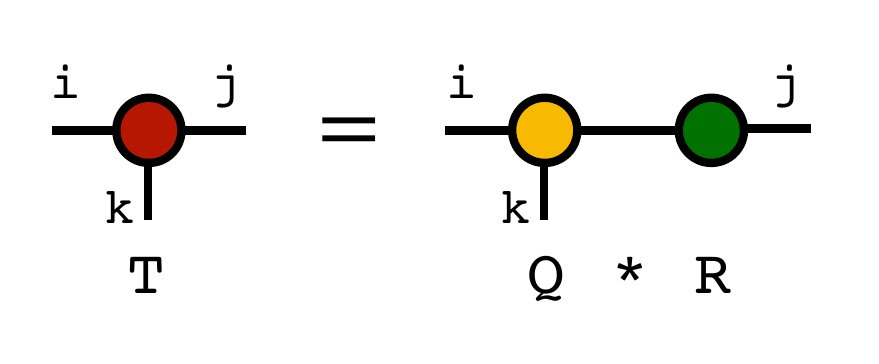}
\end{center}
A new Index is generated by the \inlinenohighlight{qr} function which
links the \inlinenohighlight{Q} tensor to the \inlinenohighlight{R} tensor as shown
above. This makes it straightforward to recover the tensor
\inlinenohighlight{T} just by using the \inlinenohighlight{*} operator:
\begin{jlcodeblock}[mathescape=true]
Q*R ≈ T  # true
\end{jlcodeblock}
(In Julia, the $\approx$ operator is overloaded to compute the relative difference
between the two sides of an equation, and return true if it is below a prescribed
threshold.)
Note that when computing the product \inlinenohighlight{Q*R} one does not 
need to know any details of the new Index introduced by the 
factorization, such as whether it is the first or second index
of R, or its dimension. However, in situations where one wants 
to retrieve this Index, a convenient way to do it is as follows:
\begin{jlcodeblock}
q = commonind(Q,R)
\end{jlcodeblock}
where the \inlinenohighlight{commonind} function returns the first Index found that
is shared by the two ITensors.

The SVD plays a key role in tensor network calculations, and is implemented as 
\begin{jlcodeblock}[mathescape=true]
W = randomITensor(i,j,m,k)
U,S,V = svd(W,(j,i))
U*S*V ≈ W  # true
\end{jlcodeblock}
\begin{center}
\includegraphics[width=0.6\columnwidth]{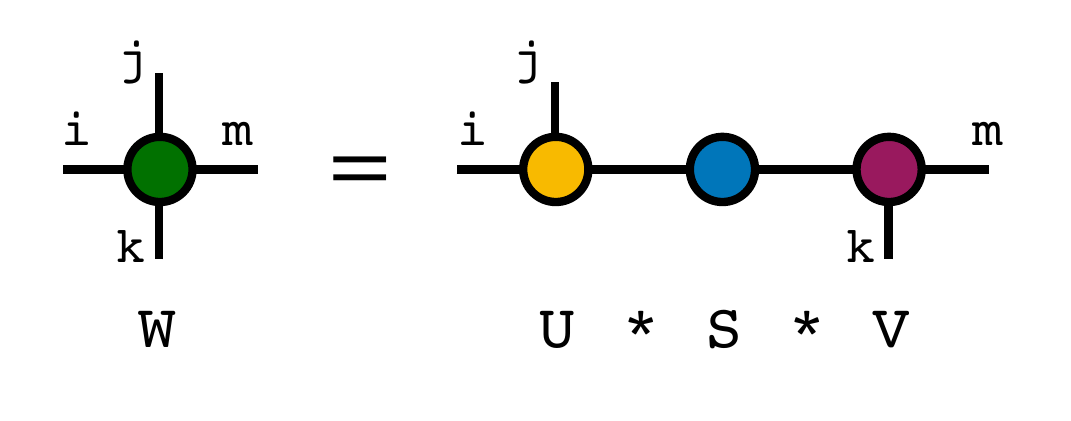}
\end{center}
In the example above, \inlinenohighlight{j,i} were specified as
the row indices, leaving \inlinenohighlight{m,k} as the column indices.

An important feature of certain decompositions such as the 
SVD is that they allow controlled truncation of the tensors
resulting from the factorization. By default, ITensor decompositions
do not truncate, though they do always compute the ``thin'' version
of a decomposition when available. A truncated decomposition
can be computed by specifying truncation keyword arguments.
In the following example
\begin{jlcodeblock}
U,S,V = svd(W,(j,i);cutoff=1E-8,maxdim=10)
\end{jlcodeblock}
the truncation will be determined by summing the squares
of the singular values from smallest to largest until the
truncation error reaches $10^{-8}$ while also ensuring
that the maximum number of singular values kept is less than or equal to 10.

\section{Tensor Storage Layer}

A powerful feature of ITensor is that ITensors can have
a wide variety of storage formats 
while offering the same user interface. 
Users can mix sparse and dense tensors together in 
calculations and manipulate any kind of tensor using
identical high-level code.

In most cases users do not set the storage type
manually; instead special storage types occur 
 automatically when using other features: 
after computing the singular value decomposition of an ITensor,
the singular values are returned as an ITensor with diagonal-sparse
storage; constructing an ITensor from indices with quantum number
subspaces makes the storage automatically block sparse.

Importantly, because the storage types used by an ITensor are 
distinct types, each one can use the most optimal memory layout possible, 
and performance-critical algorithms such as tensor
contraction and factorization can be specialized for each
storage type or combination of storage types. 
For this purpose, we take full advantage of Julia's multiple
dispatch mechanism to organize specialized algorithms into 
separate code pathways to keep the library code simple.
These optimizations are hidden from the user, who can just 
contract ITensors together using the \inlinenohighlight{*} operation and 
automatically get the best possible performance available.

Some of the most common storage types 
available in ITensor are:
\begin{itemize}
\item \emph{Dense storage}: this is the default storage type when 
constructing an ITensor from regular Index objects and setting elements.
The Dense storage type is parameterized over its element type, so 
that \inline{Dense\{Float64\}} (real-valued dense storage) and \inline{Dense\{ComplexF64\}} 
(complex-valued dense storage) are actually
different storage types. The type used to hold the data for Dense storage can also be
changed through a second, optional type parameter, to types such as \inline{Vector\{Float64\}} or \inline{SubArray\{Float64\}}.

\item \emph{Diagonal storage}: diagonal-sparse tensors occur naturally in algorithms
such as the singular value decomposition and eigenvalue decomposition. In such
settings, all of the diagonal elements can be different and so an array of the 
diagonal elements is stored. A special case of diagonal storage
is uniform diagonal storage, where all of the elements of the diagonal are constrained to
be the same. 
For this special storage only the value of the repeated, identical diagonal element is stored 
and specially-optimized contraction algorithms are invoked.
If the uniform diagonal value is equal to \inlinenohighlight{1.0} then such a diagonal tensor can 
be used to replace one Index with another under the contraction or \inlinenohighlight{*} operation, or as 
a ``copy'' or ``delta'' ($\delta$) tensor as used in certain tensor network algorithms.

\item \emph{Combiner storage}: This storage type uses essentially no memory
and stores no tensor components. Rather, it stands for a tensor which conceptually
merges two or more indices into one larger index. A combiner tensor \inlinenohighlight{C} can be created
as \inline{C = combiner(i,j,k)}
where \inlinenohighlight{i,j,k} are the indices one wants to combine together. Contracting
the combiner ITensor with an ITensor having these indices results in a new ITensor where
the indices are merged into the Index \inline{cind = combinedind(C)}.
The new combined Index is created automatically by the combiner.

\item \emph{Block sparse storage}: Block sparse storage is automatically used
when an ITensor is created from Index objects with quantum number subspaces.
This is an important case for quantum physics calculations, where the sparsity
enforces symmetries or conservation laws and allows calculations to be performed more efficiently.
The block sparse and quantum number system is discussed in more detail in 
Section~\ref{sec:qn}. 
An important consideration for block sparse storage is that the overhead of managing 
the layout of blocks and movement of blocks within algorithms must be kept very low in 
order to benefit from the efficiency of the tensor sparsity. Currently, the ITensor block sparse
storage holds all of the non-zero tensor elements in a single, contiguous array 
and keeps a dictionary mapping block indices such as \inline{(2,1,7)} to offsets in the array.

\item \emph{GPU storage}: GPU (graphics processing unit) storage is an 
experimental feature supported by the
\href{https://github.com/ITensor/ITensors.jl/tree/main/ITensorGPU}{ITensorGPU.jl} package. An ITensor
with GPU storage stores its elements in GPU memory, and calls specialized 
routines for operations including tensor contraction and tensor factorizations.
Taking advantage of the parallel processing capabilities of GPUs can give
speedups ranging from two to a hundred times the speed of CPU calculations.
Because different storage types are handled automatically behind the same
ITensor interface, GPU ITensors can take advantage of the same set of high-level
algorithms available in the ITensor library written originally for regular tensors
stored in host memory.

\item \emph{Empty Storage} \label{emptystorage}: ITensors support a special storage type \inlinenohighlight{EmptyStorage}
which is used to represent an ITensor which is numerically zero but without incurring
the cost of allocating any memory. Calling a constructor such as \inline{ITensor(i,j,k)} results
in an ITensor with empty storage. 

Another feature of the empty storage type is that it can be used as a convenient workaround for
specifying a complicated set of tensor indices in advance. A key example is when summing a set
of tensors which are known to have the same indices as each other, but where the user does not 
want or need to explicitly work with these indices. In such cases, a default-initialized ITensor (which
will have empty storage) can be used as a ``universal zero'' tensor which can be summed with any
other tensor, for example:
\begin{jlcodeblock}
i = Index(2)
V = [randomITensor(i), randomITensor(i)]
T = ITensor()
for A in V
  T += A
end
\end{jlcodeblock}
\end{itemize}

The flexibility of the ITensor storage system will let us explore other interesting 
possibilities in the future. Some planned extensions include \inlinenohighlight{IdentityStorage} storage which
represents an identity map from one collection of indices to another, \inlinenohighlight{UnitaryStorage} storage representing a unitary map,
and storage types which handle common operations such as conjugation in a lazy or delayed manner.

An important direction we plan to pursue is further sparsity patterns, including
fully general sparsity.
Technically, general sparsity is already handled by the ITensor block sparse system in the limit of all block sizes set to 1, and 
we have already observed speedups from representing sparse tensors in this limit. However, we plan to expose generally sparse
tensors more explicitly and possibly handle them in a more optimized way.
 
Lower-precision floating-point data is already supported by our storage layer, and 
can significantly speed up calculations such as when using GPU hardware.
Also we have experimental support for more exotic numerical types such as \emph{tropical} numbers, thanks to 
contributions by Jin-Guo Liu \cite{LiuTropical}.
More systematic handling of numerical types such as integer, boolean,
or nonnegative tensor elements is a planned future direction.

Given the usefulness of the flexible storage type system in ITensor, we
plan to formalize and carefully document the steps for users to make
their own custom storage types. Because of the dynamic nature of the
Julia language, such types can be fully defined outside of the ITensor
library itself yet be treated as first-class storage types for ITensors.

\section{High Level Features: MPS and MPO Algorithms}

To make ITensor a productive system for rapidly prototyping tensor network
algorithms, it provides the most common and well-developed tensor
network formats and algorithms. The two most well developed formats are
the matrix product state (MPS) tensor network \cite{Ostlund:1995,Vidal:2003,Perez-Garcia:2007}, also
known as the tensor train \cite{Oseledets:2011}, and the matrix product operator (MPO)
tensor network \cite{Verstraete:2004d,McCulloch:2007}.

Algorithms included with the core ITensor library 
include summation of MPS and MPO; truncation of MPS
and of MPO; optimization of MPS through the DMRG algorithm; 
and multiplication of an MPS by an MPO. These algorithms offer a high degree of customizability:
the multiplication of an MPS by an MPO can be performed using
at least three different algorithms (selected by a keyword argument), 
with each algorithm offering tradeoffs in terms of scaling, performance, 
and controllability. The DMRG code offers different modes, including finding the 
ground state (dominant eigenvector) of an implied sum of multiple MPOs or finding excited states (sub-dominant eigenvectors).

Throughout this section, code examples will use strings denoting local operators such as
\inlinenohighlight{"Sz"}, \inlinenohighlight{"S+"}, or \inlinenohighlight{"S-"} or strings denoting states of the local Hilbert
space such as \inlinenohighlight{"Up"} and \inlinenohighlight{"Dn"}. The way ITensor is able to know the appropriate
definition of these operators and states is through a flexible and extensible system of 
mapping operator and state names to tensors and tensor elements.

\subsection{OpSum and AutoMPO \label{sec:autompo}}

A very useful and popular feature of ITensor is the OpSum/AutoMPO system.
An OpSum is a type that lets users input
sums of products of local linear operators in a domain-specific language
and AutoMPO is the backend system for ``compiling'' these sums to MPO tensor networks.
Constructing sums of local operators is 
particularly important for physics applications, where one studies Hamiltonian
operators. A typical example being the Heisenberg Hamiltonian:
\begin{equation} \label{eq:Heis}
H = \sum_{i=1}^{N-1} \vec{S}_i \cdot \vec{S}_{i+1} = \sum_{i=1}^{N-1} S^z_i S^z_{i+1} + \frac{1}{2} S^+_i S^-_{i+1} +\frac{1}{2} S^-_i S^+_{i+1}  \ \ .
\end{equation}
This particular Hamiltonian can be exactly written as an MPO of bond dimension 5,\cite{McCulloch:2007}
but the construction is technical and tedious to program by hand.
The AutoMPO system automates the construction of this Hamiltonian MPO from the OpSum object:
\begin{jlcodeblock}
function heisenberg_mpo(N)
  # Make N S=1/2 spin indices
  sites = siteinds("S=1/2",N)

  # Input the operator terms
  os = OpSum()
  for i=1:N-1
   os +=     "Sz",i,"Sz",i+1
   os += 1/2,"S+",i,"S-",i+1
   os += 1/2,"S-",i,"S+",i+1
  end

  # Convert these terms to an MPO
  H = MPO(os,sites)

  return H
end

H = heisenberg_mpo(100)
\end{jlcodeblock}
Comparing the lines of code in the for loop above
to the Hamiltonian definition in Eq.~(\ref{eq:Heis}) one can observe a close similarity.

The AutoMPO system is powerful. Following a major enhancement
of the backend code by Anna Keselman based on Ref.~\cite{Chan:2016},
AutoMPO can accept terms with more than two local operators
and local operators separated by arbitrary distances, and
uses an SVD-based compression algorithm to obtain a nearly-optimal
MPO bond dimension.

We are working on or envision many useful extensions to this system:
\begin{itemize}
\item Improved compression techniques based on better-scaling
algorithms generalized from techniques used for non-local Hamiltonians
arising in quantum chemistry \cite{Stoudenmire:2017s}.
\item Compiling exponentials of OpSums into quantum circuits using
Trotter-Suzuki decompositions.
\item Extensions to infinite, translation-invariant systems, including
truncation methods developed for infinite MPOs like the ones
introduced in Ref.~\cite{Parker:2019}.
\item Generalizations to other tensor network topologies, such as tree
tensor networks (TTNs) and projected entangled pair operators (PEPOs).
\item Converting OpSums corresponding to interacting fermionic Hamiltonians
to free fermion approximations using mean field approximations like Hartree-Fock,
which could then be used by free fermion formulations of tensor networks 
\cite{Fishman:2015}
\footnote{\href{https://github.com/ITensor/ITensors.jl/tree/main/ITensorGaussianMPS}
{ITensorGaussianMPS.jl} is a package for constructing tensor networks of
free fermion states.}.
\end{itemize}

\subsection{DMRG Algorithm}

One of the most heavily used high-level algorithms included with ITensor is the 
density matrix renormalization group (DMRG) \cite{White:1992,Schollwoeck:2011}. 
The DMRG algorithm computes low-energy states of quantum systems, or in mathematical terms,
dominant eigenvectors of very large Hermitian linear operators. 

The main inputs to a DMRG calculation is a Hamiltonian $\hat{H}$ and an initial
guess $\Psi^{(i)}_0$ for its ground state $\Psi_0$. The ITensor DMRG implementation works 
generically for any Hamiltonian which can be represented as an MPO tensor network,
so that the same code can be applied not only to one-dimensional systems, but also
quasi-two-dimensional systems and systems with long-range interactions.
By taking advantage of the OpSum system discussed above, users can rapidly set up DMRG
calculations of complicated Hamiltonians.

Given a Hamiltonian MPO constructed as in Section~\ref{sec:autompo} above,
one can prepare an initial product state, a schedule of sweeps (DMRG algorithm iterations)
and accuracy parameters, then run the DMRG algorithm:
\begin{jlcodeblock}
# Prepare initial state MPS
state = [isodd(n) ? "Up" : "Dn" for n=1:N]
psi0_i = MPS(sites,state)

# Do 10 sweeps of DMRG, gradually
# increasing the maximum MPS 
# bond dimension
sweeps = Sweeps(10)
setmaxdim!(sweeps,10,20,100,200,400,800)
setcutoff!(sweeps,1E-8)

# Run the DMRG algorithm
energy,psi0 = dmrg(H,psi0_i,sweeps)
\end{jlcodeblock}

For Hamiltonians defined as the sum of different sets of
terms $\hat{H}=\hat{H}_1 + \hat{H}_2 + \hat{H}_3$ one
can run a DMRG calculation as:
\begin{jlcodeblock}
energy,psi0 = dmrg([H1,H2,H3],psi0_i,sweeps)
\end{jlcodeblock}
where \inlinenohighlight{H1,H2,H3} are separate MPOs. Instead of 
summing these MPOs explicitly, which can be costly and inaccurate,
the algorithm loops over them internally as if they were summed.
This technique can be helpful in applications such as quantum
chemistry where Hamiltonians can become large and complex, yet have
a nearly block diagonal MPO form if represented as a single MPO.
Expressing a Hamiltonian as a sum of MPOs also has the advantage
that parts of the DMRG algorithm, like forming the environment tensors
and diagonalizing the local effective Hamiltonian, become trivially
parallelizable\cite{Zhai:2021}.
In initial tests we found that this parallelization
is very effective and can be used in conjunction with block sparse
parallelism, which we plan to make available as a feature in future versions
of ITensor.

To compute an excited state of a Hamiltonian (sub-dominant eigenvector)
with ITensor DMRG having first computed both the ground state MPS \inlinenohighlight{psi0},
and first excited state \inlinenohighlight{psi1}, say,
one provides \inlinenohighlight{[psi0,psi1]} as an extra argument to DMRG, meaning that
the next state computed should be constrained to be orthogonal to these previous ones:
\begin{jlcodeblock}
energy,psi2 = dmrg(H,[psi0,psi1],psi2_i,sweeps)
\end{jlcodeblock}
In the implementation of this particular DMRG routine, projectors onto the
previous states \inlinenohighlight{psi0} and \inlinenohighlight{psi1} are
effectively added to the Hamiltonian times an ``energy penalty", pushing
up the energy of these states in the eigenvalue spectrum so they are no
longer part of the low-energy subspace \cite{Stoudenmire:2012a}. Other
techniques for computing excited states are planned in the future,
such as the quasiparticle MPS ansatz \cite{VanDamme, TangentSpaceMethods}.

\subsection{MPS and MPO Operations}

Far from being black-box software for performing calculations with MPS,
ITensor provides many elementary building blocks for creating custom
algorithms involving MPS, MPOs, and other tensor networks built from
these components such as projected entangled pair states (PEPS).

The most elementary interface to MPS and MPO tensor networks involves
retrieving and updating individual factor tensors making up the network.
An MPS is a factorization of a tensor \inlinenohighlight{psi} of the following form
\begin{equation}
\psi^{s_1 s_2 \cdots s_N} = \sum_{\{\alpha\}} A^{s_1}_{\alpha_1} A^{s_2}_{\alpha_1 \alpha_2} A^{s_3}_{\alpha_2 \alpha_3} \cdots A^{s_N}_{\alpha_N} \ ,
\end{equation}
where we have omitted an explicit site-label $j$ on each $A$ tensor for compactness.
The factor tensor $A^{s_j}_{\alpha_{j-1} \alpha_j}$ on site $j$ can be obtained as
\begin{jlcodeblock}
A = psi[j]
\end{jlcodeblock}
and updated as
\begin{jlcodeblock}
psi[j] = new_A
\end{jlcodeblock}

To analyze the properties of an MPS, one is often interested in expected values of
local operators.
To compute the expected value of an operator at every site and return an array of the results,
one can use the function \inlinenohighlight{expect}. For example, calling
\begin{jlcodeblock}
avgSz = expect(psi,"Sz")
\end{jlcodeblock}
on an MPS \inlinenohighlight{psi} will compute $\bra{\psi}\hat{S}^z_j\ket{\psi}$ for every site
$j$ and return an array of the results, where here we use the example of the spin $\hat{S}^z$ operator as our local operator.

Another common quantity of interest is the two-point correlation function of a pair of
local operators acting at distant sites $i$ and $j$. Using the example of a spin system
again, let us say we are interested in the correlation matrix given by $C_{ij} = \bra{\psi}\hat{S}^+_i \hat{S}^-_j\ket{\psi}$. This correlation matrix can be efficiently computed as:
\begin{jlcodeblock}
C = correlation_matrix(psi,"S+","S-")
\end{jlcodeblock}
The \inlinenohighlight{correlation_matrix} function accepts optional keyword arguments such as a smaller range of sites
over which to compute the correlation matrix, versus the whole system. It also automatically ensures correct results for fermionic operators such as \inlinenohighlight{"Cdag"} and \inlinenohighlight{"C"} (spinless fermion $\hat{c}$ and $\hat{c}^\dagger$ operators). 

An important technical step involving an MPS is bringing it into an
orthogonal form, where all of the factor tensors to the left or right of
the \emph{center tensor} at a site $j$ are equivalent to partial isometries (i.e. either their rows or their columns are orthogonal).
To bring an MPS into orthogonal form efficiently in ITensor, one calls:
\begin{jlcodeblock}
orthogonalize!(psi,j)
\end{jlcodeblock} 
where we follow the convention adopted in Julia programming that functions whose
name end with \inlinenohighlight{!} may modify their first argument.
An interesting feature of ITensor MPS objects is that they store information about
which tensors are known to be orthogonal, so that calling
\inline{orthogonalize!(psi,j)} repeatedly for the same value of \inlinenohighlight{j} 
does no extra work, and shifting
the orthogonality center of an already partially orthogonalized MPS 
can be done with the minimum amount of computation.

Another fundamental operation is truncating an MPS: computing another MPS of
a smaller bond dimension which is as close to the original MPS as possible.
For MPS such a truncation can be done optimally through various deterministic algorithms.
Truncating an MPS \inlinenohighlight{psi} in ITensor can be done by calling:
\begin{jlcodeblock}
truncate!(psi;maxdim=500,cutoff=1E-8)
\end{jlcodeblock} 
where for the sake of example we have shown specific values of the two most commonly
used truncation parameters. The \inlinenohighlight{maxdim} parameter sets an upper limit on the
bond dimension of the MPS after the truncation, whereas the \inlinenohighlight{cutoff} parameter
allows the new bond dimension to be determined adaptively as long as the resulting
truncation error remains below the value provided. Using a cutoff can allow the
bond dimension to fall below the \inlinenohighlight{maxdim} when possible while still ensuring an accurate 
approximation of the original MPS.

ITensor supports arithmetic involving MPS and MPOs to be performed using the 
\inlinenohighlight{add} function. Performing exact sums can lead to quickly growing costs,
so that one normally truncates the result by providing a truncation-error cutoff.
For example, to add two MPS \inlinenohighlight{psi} and \inlinenohighlight{phi} one can call:
\begin{jlcodeblock}
eta = add(psi,phi;cutoff=1E-10)
\end{jlcodeblock}
and similarly for adding two MPOs. Currently this method uses a particular backend algorithm
known as the ``density matrix'' algorithm \cite{densitymatrix} 
but other backends will be available in the future to select through an optional keyword argument.

Algorithms such as time-evolving quantum states or contracting two-dimensional ``PEPS'' tensor 
networks can be formulated in terms of products of an MPO with MPS or with another MPO.
To approximately multiply an MPS \inlinenohighlight{psi} by an MPO \inlinenohighlight{W}, one can call the function
\begin{jlcodeblock}
Wpsi = contract(W,psi;maxdim=50)
\end{jlcodeblock} 
with example parameters controlling the truncation shown.
The product of two MPOs \inlinenohighlight{R} and \inlinenohighlight{W} can also be computed:
\begin{jlcodeblock}
RW = contract(R,W;cutoff=1E-9)
\end{jlcodeblock} 
Importantly, these functions provide multiple backend algorithm implementations 
with various tradeoffs in terms of the cost, accuracy, and control offered.
For example, to select the accurate yet expensive ``naive'' algorithm 
for multiplying an MPS by an MPO one may call
\begin{jlcodeblock}
Wpsi = contract(W,psi;method="naive")
\end{jlcodeblock}

\section{Quantum Number Block Sparse ITensors \label{sec:qn}}

An important technique used in state-of-the-art physics calculations
is enforcing constraints on tensors arising from conserved quantities. 
These are quantities such as total particle number or total spin along an axis
which are conserved due to symmetries of the Hamiltonian operator.
The value of each conserved quantity is known as a \emph{quantum number}.

Quantum number conservation can be important since
physical systems commonly respect symmetries such as rotational symmetry 
or particle number conservation symmetry, making it necessary for simulations to conserve these
 to be comparable to experimental results. 
Just as importantly, conserving quantum numbers allows calculations to run much faster
and use less memory because of a \emph{block sparse} structure that is naturally 
imposed on the tensors in a tensor network \cite{SinghU1}. A detailed discussion of
structures imposed by symmetries on tensors and tensor networks is given in Refs. \cite{SinghGlobal,SinghU1,SinghSU2}.

The power of the ITensor approach to conserving quantum numbers is that quantum number
conserving ITensors offer nearly the same interface as regular, dense ITensors.
Algorithms can be written generically for dense ITensors and automatically
work for the symmetric case too, as long as tensors are correctly conjugated using the \inlinenohighlight{dag}
function, which would be necessary to use to obtain correct results with complex-valued
tensors anyway.

The design of the ITensor quantum number (QN) system is that QN information is stored
in Index objects in a fixed order. This information is queried when an ITensor is constructed to determine
whether the storage should be block sparse, as well as the layout of the blocks, and which blocks are allocated. 
When such ITensors are summed, contracted, or factorized, optimized 
routines are used and the QN information is propagated to the indices of the resulting ITensor.

Currently ITensor only supports quantum numbers arising from symmetries under Abelian groups such as
$U(1)$ or $\mathbb{Z}_n$, which are ubiquitous in physics. We are also in the planning stages of support for 
non-Abelian symmetries such as $SU(2)$ in a future version of ITensor, but the remainder of this section
will discuss only the Abelian case.

As an illustrative example of ITensor's QN system, 
say we have defined two indices with information about their QN subspaces:
\begin{jlcodeblock}
i = Index(QN(0)=>2,QN(1)=>3;tags="i")
j = Index(QN(1)=>2,QN(2)=>1;tags="j")
\end{jlcodeblock} 
The Index \inlinenohighlight{i} has a total dimension of 5 because it has two subspaces, one carrying a quantum
number \inline{QN(0)} and of dimension of 2; the other carrying a quantum number \inline{QN(1)} 
and of dimension 3. Similarly \inlinenohighlight{j} has a total dimension of 3, coming from its two subspaces.

Using these indices, we can define an ITensor \inlinenohighlight{T} in the usual way as
\begin{jlcodeblock}
T = ITensor(i,j)
\end{jlcodeblock} 
where initially this ITensor will have Empty storage (see Sec.~\ref{emptystorage}), and thus an as-yet unspecified pattern of non-zero blocks.
Then, we set an element of \inlinenohighlight{T} as
\begin{jlcodeblock}
T[i=>3,j=>1] = 31.0
\end{jlcodeblock}
Note that this element corresponds to the \inline{QN(1)} subspace of \inlinenohighlight{i} and the \inline{QN(1)}
subspace of \inlinenohighlight{j}, for a combined ``QN flux'' of \inline{flux(T) == QN(2)}. (Mathematically the flux 
corresponds to the overall irreducible representation under which the tensor transforms. More intuitively, it describes
whether a tensor is a source or sink of quantum numbers and by how much.)
When setting any further elements, only those elements of \inlinenohighlight{T} consistent with a 
flux \inline{QN(2)} will be allowed to be non-zero.
This constraint imposes a block-sparse structure on \inlinenohighlight{T}, since most values of the indices combine
to form fluxes other than \inline{QN(2)} and thus remain zero. Only allowed blocks consistent
with the total flux are stored in memory. Block-sparse computations can then be much more 
efficient than with dense tensors because fewer non-zero elements have to be handled and 
the presence of disjoint blocks allows major parts of calculations to be performed in parallel.

For the rest of this section, we discuss in more detail the different 
types composing the QN block sparse ITensor system.

\subsection{QN Objects}

Block-sparse ITensors arise from vector spaces which are a direct sum of smaller
subspaces. In physics calculations, these subspaces are associated with different 
\emph{quantum numbers}. In ITensor, sets of quantum numbers are stored in QN objects
as a collection of name-value pairs, where the value is always an integer. Different 
values may be combined according to the usual rules of integer addition and subtraction, 
possibly modulo some other integer $N$. (For the case of quantum numbers arising from
 non-Abelian symmetries, these rules must be generalized.)

QN objects carrying a single quantum number, such as total z-component spin \inlinenohighlight{"Sz"}, may
be constructed as:
\begin{jlcodeblock}
q0 = QN("Sz",0)
q1 = QN("Sz",1)
\end{jlcodeblock} 
QNs may be added, subtracted, and compared:
\begin{jlcodeblock}
q0 + q1 == QN("Sz",1) # true
q1 + q1 == QN("Sz",2) # true
\end{jlcodeblock}

QN objects can also carry multiple quantum numbers as follows:
\begin{jlcodeblock}
a = QN(("N",0),("Sz",0))
b = QN(("N",1),("Sz",-1))
\end{jlcodeblock} 
Because the quantum numbers are named, they can be provided to the QN constructor in any order and are 
sorted internally.
For convenience when there is only one quantum number, its name can be omitted; this is 
equivalent to choosing the name to be the empty string.

Some quantum numbers of physical systems obey a $\mathbb{Z}_N$ addition rule. A key example
is fermion parity, which is only conserved modulo two in systems such as superconductors.
A $\mathbb{Z}_N$ addition rule for a quantum number can be specified by providing $N$ as
the third entry of the tuple defining that quantum number:
\begin{jlcodeblock}
p0 = QN("P",0,2)
p1 = QN("P",1,2)
p1 + p1 == QN("P",0,2)
\end{jlcodeblock}
The \inlinenohighlight{2} following the quantum number values above specifies that the \inlinenohighlight{"P"} quantum
number obeys $\mathbb{Z}_2$ addition.

The reason quantum numbers have names and are not just distinguished positionally
is that having names allows QNs containing different quantum numbers to be combined automatically
and correctly.
This becomes important when different local physical spaces (such as 
spin versus particle degrees of freedom) are defined separately, then combined or mixed later.
Key examples of physical models combining two otherwise separate types of physical spaces
are the Hubbard-Holstein model, where electron sites are intermixed with boson sites, or
the Kondo model mixing electron sites with spin sites.

\subsection{QN Index}

As discussed above, the block-sparse structure of quantum number 
conserving tensors arises from the direct-sum structure of the 
vector spaces over which they are defined. To specify additional
information about direct-sum subspaces, an \inlinenohighlight{Index} object
can be constructed from QN-integer pairs, as follows:
\begin{jlcodeblock}
i = Index(QN("N",0)=>1,
          QN("N",1)=>3,
          QN("N",2)=>2; tags="i")
\end{jlcodeblock}
where we note that \inline{(a=>b) == Pair(a,b)} is built-in Julia notation for 
constructing a pair of values \inlinenohighlight{a} and \inlinenohighlight{b}. 

In the example above, the Index \inlinenohighlight{i} has three subspaces,
of dimensions 1, 3, and 2 respectively. Therefore the total dimension
of \inlinenohighlight{i} is six, or \inlinenohighlight{dim(i) == 6}.
The subspaces are associated with the quantum numbers \inline{QN("N",0)},
\inline{QN("N",1)}, and \inline{QN("N",2)} respectively.

A crucial aspect of QN Index objects not yet discussed is that they have
an \inlinenohighlight{Arrow} direction, which can be \inlinenohighlight{Out} or \inlinenohighlight{In}, with \inlinenohighlight{Out}
being the default. Mathematically, the direction of an Index says whether it is
covariant (\inlinenohighlight{In}) or contravariant (\inlinenohighlight{Out}) and expresses how the Index
transforms under the symmetry group action. A  physicist might view an \inlinenohighlight{Out} arrow as
denoting a ``ket'' index and an \inlinenohighlight{In} arrow as a ``bra'' index.
The arrows of QN indices play two important roles in working with QN ITensors:
\begin{itemize}
\item A pair of QN indices must have \emph{opposite} arrow directions to be contracted.
\item When computing the QN flux of an ITensor block, QNs corresponding to an \inlinenohighlight{Out}
Index are added and QNs corresponding to an \inlinenohighlight{In} Index are subtracted.
\end{itemize}
Examples of these arrow and flux rules will be given in the next section on QN ITensors.

\subsection{QN ITensor}

Constructing an ITensor from QN Indices makes it a QN ITensor, with a block sparse storage
type. In addition to the block sparse real and complex storage types, there are also diagonal
block sparse storage types which are usually obtained from factorizations such as the 
SVD of block sparse ITensors.

In most respects, working with QN ITensors is quite similar to working with dense ITensors.
Operations like adding QN ITensors or multiplying them by scalars work
in a straightforward way.
However, one small but important difference from dense ITensors arises when contracting QN ITensors:
matching QN indices must have opposite arrow directions to be contracted. This rule is important for 
consistent bookkeeping of QN flux under Hermitian conjugation of ITensors. 
But because computing the 
Hermitian conjugate \inline{dag(T)} of a QN ITensor \inlinenohighlight{T} is defined to reverse all of the 
arrows of its indices, code which is already written correctly for complex, dense ITensors 
(with proper use of \inlinenohighlight{dag} to handle complex conjugation) will automatically be correct
in terms of QN conservation too.


Having discussed all of the types involved in the QN ITensor system, let us discuss some 
examples which integrate all of these elements. An example motivated by physics is the
Hilbert space of a single ``hard-core'' boson: a type of particle which cannot share an
orbital or site with another boson. Such bosons can be used to model atoms which have large, 
short-range repulsive interactions. The Hilbert space of a single hard-core
boson is spanned by two basis states $\ket{0}$ and $\ket{1}$, representing no particle and
one particle. Along with these basis states, one can define the elementary operators 
$a$, $a^\dagger$, $n$, which lower, raise, or measure the number of particles:
\begin{align}
a \ket{1} & = \ket{0} \nonumber \\
a^\dagger \ket{0} & = \ket{1} \nonumber \\
n \ket{0} & = 0 \nonumber \\
n \ket{1} & = \ket{1}
\end{align}
Diagrammatically the equation $a \ket{1} = \ket{0}$ can be expressed as
\begin{center}
\includegraphics[width=0.3\columnwidth]{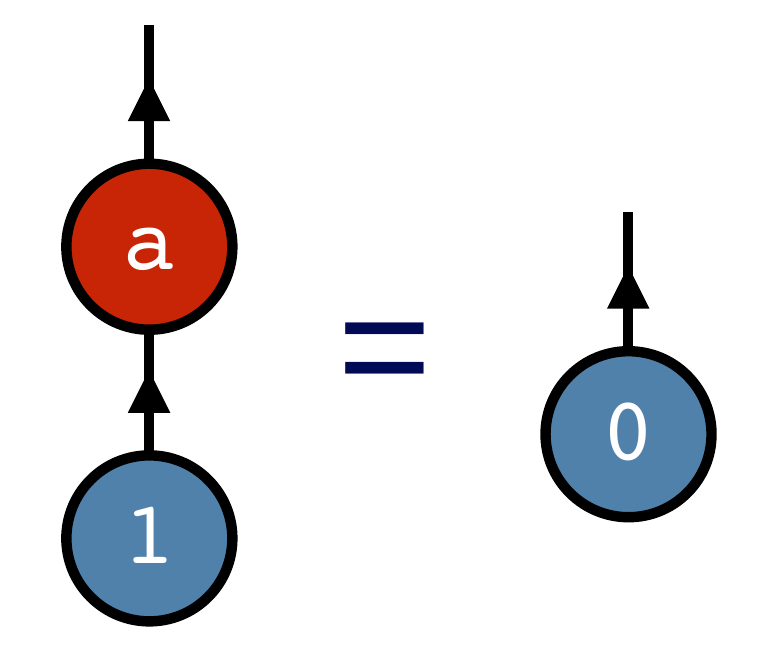}
\end{center}
where in the diagram note that the tensors now have arrows on their indices,
with contracted indices having opposite arrow directions (\inlinenohighlight{In} versus \inlinenohighlight{Out}).
Within ITensor, we can represent the Hilbert space of this boson as an \inlinenohighlight{Index}
\begin{jlcodeblock}
s = Index(QN("N",0)=>1,
          QN("N",1)=>1;
          tags="Boson")
\end{jlcodeblock}
This \inlinenohighlight{Index} is the representation in code of the index lines \includegraphics[width=0.03\columnwidth]{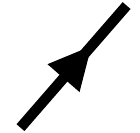}
in the $a \ket{1} = \ket{0}$ diagram above. By default, Index objects have an \inlinenohighlight{Out} arrow direction meaning a contravariant index.

We can next construct the operator $a$ as the following ITensor
\begin{jlcodeblock}
a = ITensor(s',dag(s))
a[s'=>1,s=>2] = 1.0
\end{jlcodeblock}
The first line constructs \inlinenohighlight{a} as an ITensor with indices \inline{s'} and \inline{dag(s)}
 with elements all zero, and the second
line sets the only non-zero element of \inlinenohighlight{a}. We can visualize the resulting tensor
as follows
\begin{center}
\includegraphics[width=0.45\columnwidth]{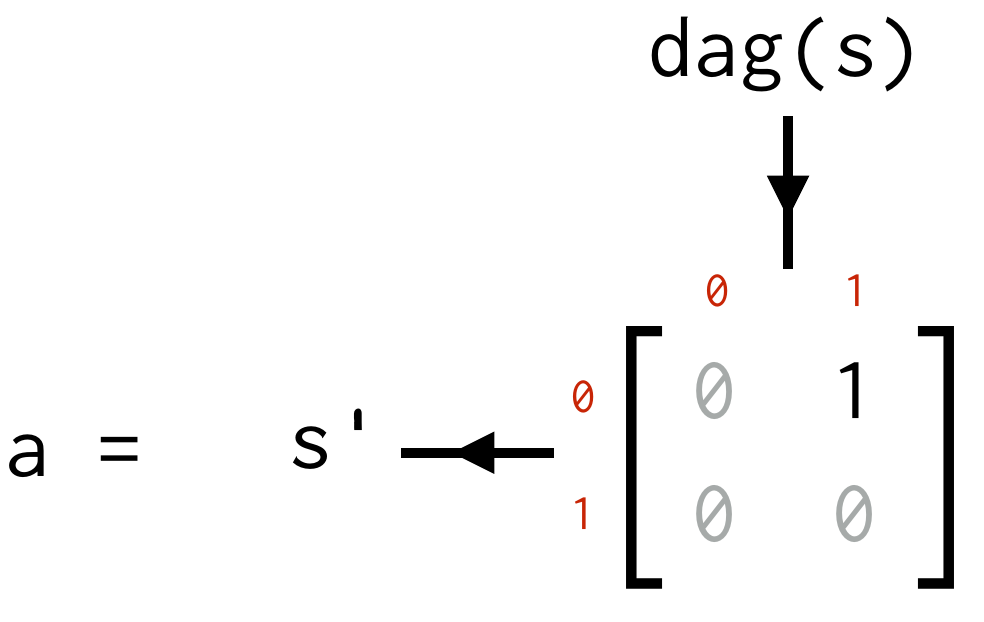}
\end{center}
The small, red labels above denote the subspaces of the Index \inlinenohighlight{s}
by labelling them according to the value of the \inlinenohighlight{"N"} quantum number.
We can also see the single non-zero element corresponding to the (1,2) entry
of the tensor and having the value 1.0. 

A key point about the example of the ITensor for the $a$ operator is
 that \emph{the only element stored in memory is the one shown
above}. All other entries shown in light gray are assumed zero and not stored in memory.
To see why this is the case, let us label each block of the $a$ tensor (or any tensor 
having the same indices as $a$) by its quantum number flux:
\begin{center}
\includegraphics[width=0.3\columnwidth]{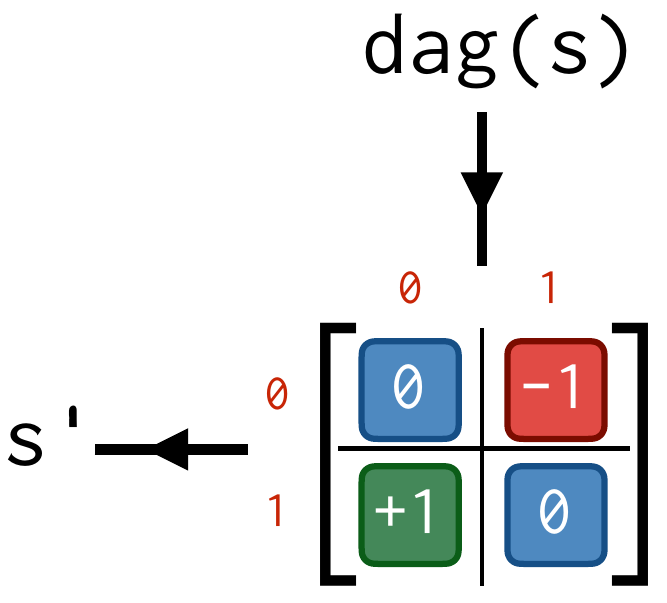}
\end{center}
The non-zero element of the tensor $a$ is in the block with flux \inline{QN("N",-1)} and physically
means this operator always reduces the particle number by 1. Because the convention in ITensor
is that QN-conserving ITensors must have a well-defined flux, only blocks with the same flux 
are stored in memory and the rest are assumed to be zero and not stored.
In contrast, the $n$ operator has a flux of zero, and therefore will have
two allowed blocks: the blocks labeled $0$ and shown in blue in the diagram above.

Unlike the examples above, general QN-conserving ITensors will have many blocks which
can be non-zero, and having block sizes greater than $1\!\times\!1$. Summations, factorizations,
and especially contractions of general block sparse tensors can be much faster than for dense tensors with the same index dimensions, not only because the zero (unallocated) blocks can be skipped over,
but also because operations on non-zero blocks can be performed in parallel. Both the Julia and C++ implementations of ITensor use multi-core parallelism within their block sparse tensor
contraction algorithm, with speedups of up to $5\times$ observed in practical physics applications, though the speedups vary depending on the application.

Finally, all other operations available for dense tensors work for QN-conserving ITensors too, with
exactly the same interface. This includes the use of combiner ITensors, factorizations such as the SVD
and QR, and higher-level algorithms involving matrix product states and operators. 
For further reading on how various tensor operations can be implemented while respecting Abelian group symmetries and related quantum
numbers, see Ref.~\cite{SinghU1}.

\section{NDTensors Library}

Early on in the design of the ITensor library, a conscious decision was made
to separate the high-level ITensor interface, involving ``intelligent'' Index
objects and related features, from the lower-level parts of the code focusing on 
efficient contraction routines and sparse tensor storage layouts.
With the port of the ITensor library to Julia, we have taken this design one 
step further by making the lower-level part of the library a separate
submodule\footnote{By a module and a submodule we mean a separate namespace for defining types and methods.} known as \href{https://github.com/ITensor/ITensors.jl/tree/main/NDTensors}{NDTensors} ($N$-Dimensional Tensors) which
can be used and developed separately from ITensors.jl\footnote{At the time of
writing this paper, the NDTensors library can only
be installed by installing the ITensors library for convenience of developing
the libraries in tandem, however we plan to split it off so it can be
installed seperately from ITensors in the near future.}.

Some of the goals of developing NDTensors as a separate module include:
\begin{itemize}
\item Separating low-level NDTensors algorithms from high-level ITensor logic simplifies and modularizes the code and prevents bugs.
\item Encouraging more community contributions to the ITensor project, since some community
members may find the NDTensors interface and features more familiar and appealing, and
may not prefer to work with the ITensor layer when making contributions.
\item Other software besides ITensor could eventually use NDTensors as a backend, which would promote community efforts to improve tensor software and share resources. Fully realizing this possibility would require releasing it as a separate library in the future, which we plan to do.
\end{itemize}

\subsection{Basic Interface}

The NDTensors library is a full-featured, standalone library emphasizing generic, high-performance
algorithms and support for a variety of sparse tensor types. Unlike the ITensor library, 
NDTensors has a more traditional interface where users must keep track of the ordering
of tensor indices. For example, one can construct a dense tensor with dimensions $3,7,4$ as
\begin{jlcodeblock}
using ITensors.NDTensors

T = Tensor(3,7,4)
\end{jlcodeblock}
which by default is filled with all zeros and then set its elements as
\begin{jlcodeblock}
T[1,2,1] = 1.23
T[3,2,3] = -0.456
\end{jlcodeblock}
Tensor objects are 1-indexed, similar to Julia arrays.
A tensor with complex entries can be constructed as
\begin{jlcodeblock}
T = Tensor(ComplexF64,5,4,3)
\end{jlcodeblock}

Contracting two Tensors is done by specifying
temporary labels for tensor indices; matching labels indicate two indices
are contracted while unique labels denote uncontracted indices.
In the following example:
\begin{jlcodeblock}
A = randomTensor(3,7,2)
B = randomTensor(4,2,3)
C = contract(A,(-1,1,-2),B,(2,-2,-1))
\end{jlcodeblock}
the label \inlinenohighlight{-1} of the first index of \inlinenohighlight{A} matches the \inlinenohighlight{-1}
label of the third index of \inlinenohighlight{B}, so those two indices are contracted with each 
other. Likewise the third index of \inlinenohighlight{A} and second of \inlinenohighlight{B} share
the label \inlinenohighlight{-2} and are contracted. The use of negative integers to label contracted indices 
is not required, but is just a convention to make the code more readable.

\subsection{Block Sparse Tensors}

NDTensors provides sparse tensor types as well. An important example is block
sparsity. One way to construct a block sparse tensor is as follows:
\begin{jlcodeblock}
blockdims = ([2,2],[2,3])
nzblocks = [(1,2),(2,1)]
A = randomBlockSparseTensor(nzblocks,blockdims)
\end{jlcodeblock}
The code above specifies that the tensor \inlinenohighlight{A} has two indices of dimension \mbox{4 (= 2+2)} and 
\mbox{5 (= 2+3)} respectively, with the first index having two subspaces of dimensions 2 and 2 and the 
second index having two subspaces of dimensions 2 and 3. Thus \inlinenohighlight{A} has four blocks overall, 
because its two indices each have two subspaces. The array \inlinenohighlight{nzblocks} lists which
blocks of \inlinenohighlight{A} can be non-zero and will be actually allocated in memory, with each tuple giving a subspace number for each index.
We can visualize a typical result for the tensor \inlinenohighlight{A} as follows:
\begin{center}
\includegraphics[width=0.6\columnwidth]{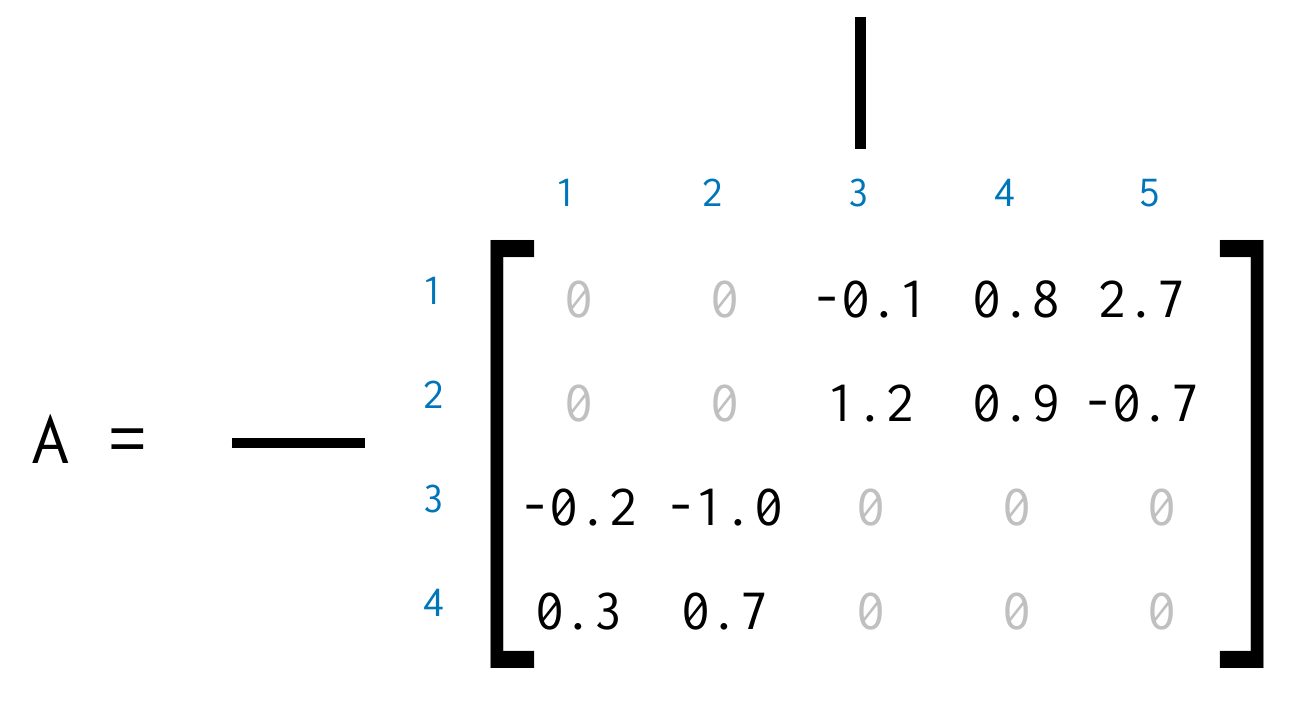}
\end{center}
where the zeros shown in light gray are only assumed and not actually allocated in memory.

\subsection{Generic Index Types}

A crucial feature of NDTensor is that tensors are allowed to represent their indices not just
as a collection of integers or block dimensions (specifying each the dimension of each index), 
but as any object providing a certain index interface. 
This generic design allows seamless interoperation between the NDTensors library and
the ITensor library, as well as making it easy to provide features such as tensor slicing. 

For dense and diag storage, essentially all that is required of the container \inlinenohighlight{inds} representing the indices of a \inlinenohighlight{Tensor}
is that one can call the function \inlinenohighlight{dim} on its \inlinenohighlight{n}th element. 
Examples of valid \inlinenohighlight{inds} objects are collections of integers, collections of ITensor \inlinenohighlight{Index} objects (provided an overload of the method \inlinenohighlight{dim} is provided), \inlinenohighlight{Dims} objects 
provided by the Julia Base library for indexing built-in Julia tensors, 
and \inlinenohighlight{BlockDims} objects defined by NDTensors for indexing block sparse tensors.
By default, the strides are determined by the dimensions
of the indices, but can be overloaded if needed such as for tensor slicing applications.
Unless an explicit set of indices is provided, \inlinenohighlight{Tensor} objects default to using the \inlinenohighlight{Dims}
type (a tuple of integers) to represent its indices and \inlinenohighlight{BlockSparseTensor} objects default to using \inlinenohighlight{BlockDims}.
For block sparse storage types, an overload of the \inlinenohighlight{blockdim} function is required for any block index, which is used to query the size of a specified block in a specified dimension.

\subsection{Tensor Contraction Backend}

Tensor contractions are often the computational bottleneck of tensor network algorithms. Thus 
implementing it as efficiently as possible is critical for performance.

For contracting two dense tensors, NDTensors currently uses a strategy of permuting and reshaping the
tensors into matrices, so that the contraction maps to a matrix multiplication\footnote{This is sometimes referred to as the Transpose-Transpose-GEMM-Transpose (TTGT)\cite{TCL} approach}. The motivation
behind this strategy is that BLAS libraries such as Intel MKL offer such high performance that
the extra overhead of permuting the tensors is worthwhile. It is also important to note that 
the tensor permutation has a sub-leading scaling relative to the matrix multiplication, so that in
the limit of large tensors the computation is dominated by the BLAS \inlinenohighlight{dgemm} or \inlinenohighlight{zgemm} routines. 
Though this strategy is a common one for tensor libraries, its implementation in NDTensors is done
 carefully to ensure that every case where permutation can be avoided is taken advantage of. Also
if two equivalent strategies exist to permute the contracted tensors to matrices where one of the 
permutations is trivial, the code chooses to permute the smaller of the two tensors.

The case of block sparse tensor contraction reduces to doing a set of smaller, dense tensor contractions
on various pairs of blocks from the tensors being contracted
\footnote{Currently the default in ITensor is that block sparse tensors are contracted directly
without first reshaping into a matrix. An alternative is to first permute and reshape the block sparse
tensors into block sparse matrices. With that strategy, degenerate quantum number blocks can be combined,
leading to a contraction involving a smaller number of larger blocks, which is advantageous for BLAS 
\cite{Evenbly:PrivateCommunication}.
This alternative contraction strategy can be enabled with the experimental \inline{ITensors.enable_combine_contract()} 
function which enables a global flag.
Currently we find that neither of the two strategies (contracting versus combining then contracting)
is better in every situations, and it depends on details like the quantum numbers, sparsity and order of the
tensors being contracted.}.
Thus it is built on top of the dense
contraction layer of NDTensors, but also offers an excellent opportunity to exploit parallelism, since
contraction of the blocks can be done independently, although one does have to handle cases where multiple
block pairs contribute to the same block of the resulting tensor.
By exploiting multi-core parallelism for the same algorithm within the C++ implementation of ITensor 
we have observed speedups of $2-3\times$ for DMRG and related MPS calculations (depending on the sparsity
and block sizes involved, which varies strongly based on the symmetries used), and up to $5\times$ for
tree tensor network calculations.
More recently, we have implemented the same kind of multi-core parallelism in the block sparse contraction
algorithm in NDTensors using Julia's native multithreading and have seen similar speedups to those we
see in the C++ implementation that uses OpenMP.

Looking ahead, a key improvement to NDTensors will be to offer support for more advanced tensor contraction
algorithms that have been recently developed. These algorithms build on sophisticated research into BLAS
software, where it was realized that modern BLAS implementations could apply to the case of tensors of
arbitrary order, and not just matrices. The two implementations of this type we are aware of are 
TBLIS \cite{TBLIS} and TCL/GETT \cite{TCL}. These libraries significantly reduce, if not totally eliminate,
the permutation overhead inherent to the permute-to-matrix strategy discussed above, offering superior performance
to the default contraction algorithm of C++ ITensor and NDTensors 
\cite{benchmarks}.
We currently have an experimental feature in the Julia version of ITensors.jl that provides TBLIS
as an optional contraction backend, and have seen speedups over our current contraction code, particularly
when contracting larger tensors. We plan to do benchmarks using this TBLIS
backend for more sophisticated algorithms like DMRG in the near future.

\section{Other Features of ITensor \label{sec:features}}

The ITensor library has many other features which are important for productive programming, developing new algorithms or treating new problem domains, but whose precise details are somewhat beyond this high-level introduction. In this section we briefly highlight these features.

\subsection{Writing and Reading ITensor Objects with the HDF5 Format}

One important feature is that nearly every type involved in a tensor network, from Index objects to IndexSet's to ITensors, MPS, and MPOs can be written to and read from HDF5 files. The HDF5 format is a widely used and standardized format for writing large datasets and heterogenous data. It offers portability across operating systems with different binary formats; metadata and a file-system structure for organizing and retrieving data; and efficient use of memory including compression of numerical data. ITensor objects written to HDF5 files can be both written to and read from both the C++ and Julia versions of ITensor, allowing users with large C++ codes to use the new Julia version for tasks such as performing analysis of simulation results.

\subsection{Defining Custom Local Hilbert Spaces \label{sec:custom_hilbert}}

\begin{jlcodeblock}[frame=single,float,caption={Overloads of the $\mathtt{ITensors.op}$ method which define custom mappings of operator names to ITensors for Index objects having the tag "S=3/2".},captionpos=b,label=lst_op]
using ITensors
import ITensors: op  #allows overloading of ITensors.op

op(::OpName"Sz",::SiteType"S=3/2") = [
  +3/2   0    0    0
    0   +1/2  0    0
    0    0   -1/2  0
    0    0    0   -3/2 
] 

op(::OpName"S+",::SiteType"S=3/2") = [
    0  sqrt(3) 0    0
    0    0     2    0
    0    0     0  sqrt(3)
    0    0     0    0
] 

op(::OpName"S-",::SiteType"S=3/2") = [
    0    0    0     0
 sqrt(3) 0    0     0
    0    2    0     0
    0    0  sqrt(3) 0
] 

\end{jlcodeblock}

An important feature for physics applications is the ability to define custom ``degrees of freedom'' or local Hilbert spaces and associated local operators to allow users to implement their own systems of interest within high-level tools like OpSum. ITensor includes built-in definitions for only a handful of common cases such as $S=1/2$ and $S=1$ spin degrees of freedom, spinless and spinful fermions, and the Hilbert space of the $t\!-\!J$ model. But physics applications of ITensor often call for other definitions, such as of local Hilbert spaces for bosons, higher spin moments such as $S=3/2$, and more exotic degrees of freedom such as $\mathbb{Z}_N$ parafermions. Users may also want to extend built-in Hilbert space types by defining additional local operators. The C++ version of ITensor already lets users define custom local Hilbert spaces and operators, but due to limitations of the C++ language the customization process has remained cumbersome and users have often had trouble mastering the necessary tasks of defining C++ types, constructors, and methods.

Fortunately, in the Julia version of ITensor we have been able to streamline the process of defining and using custom Hilbert spaces. The key innovation is that certain Index tags can be designated as special by defining associated ``site types''. For example, say a user wants any Index carrying the tag \inlinenohighlight{"S=3/2"} to be interpreted as a $S=3/2$ spin (the Index should also have the appropriate dimension of 4). Practically this means we want systems such as OpSum to know how to make the appropriate local operators such as \inlinenohighlight{"Sz"}, \inlinenohighlight{"S+"}, and \inlinenohighlight{"S-"} which act on the Hilbert space of this Index. To tell ITensor how these operators should be defined, a user can create overloads of the \inlinenohighlight{ITensors.op!} method which accept a special type: \inline{SiteType"S=3/2"}. Examples of such overloads are shown in Listing~\ref{lst_op} and can be defined outside the ITensor library in user code.
The notation \inline{SiteType"S=3/2"} is a convenient Julia macro syntax which is used to create a 
unique type parameterized by a string. Creating types out of values 
allows one to effectively overload functions over 
different values, even though technically functions can only be overloaded over different types.

After defining these functions, the following code will return \inlinenohighlight{"Sz"}, \inlinenohighlight{"S+"}, and \inlinenohighlight{"S-"} operators as ITensors given an Index \inlinenohighlight{s} which has the \inlinenohighlight{"S=3/2"} tag
\begin{jlcodeblock}
s = Index(4,"S=3/2") # make an Index with the tag "S=3/2"
Sz = op("Sz",s)
Sp = op("S+",s)
Sm = op("S-",s)
\end{jlcodeblock}
The ITensor library reads the tags of the Index passed as the second argument to \inlinenohighlight{op}, then checks if any of these tags have an associated \inlinenohighlight{SiteType} overload of \inlinenohighlight{ITensors.op}. If exactly one tag and operator name pair does have an \inlinenohighlight{ITensors.op} method defined for it, such as the  \inline{::SiteType"S=3/2"}, \inline{::OpName"Sz"}  overload in Listing~\ref{lst_op} above, then that overload is called to produce the operator corresponding to the requested name as an ITensor. Users can also overload other functions which both construct and return the operator ITensor, giving more control over the whole process.

What makes this system powerful is that the same \inlinenohighlight{op} method and its overloads are called by the OpSum system and various MPS and MPO constructors within ITensor library code. So after defining the \inline{SiteType"S=3/2"} overloads of the \inlinenohighlight{op!} or \inlinenohighlight{op} functions above, the following code ``just works'' and correctly makes an MPO of the Heisenberg Hamiltonian for an $N$-site system of $S=3/2$ spins:
\begin{jlcodeblock}
sites = [Index(4,"S=3/2,n=$n") for n=1:N]
         
os = OpSum()
for j=1:N-1
  os +=     "Sz",j,"Sz",j+1
  os += 1/2,"S+",j,"S-",j+1
  os += 1/2,"S-",j,"S+",j+1
end

H = MPO(os,sites)
\end{jlcodeblock}
Various special tags with associated \inlinenohighlight{SiteType} operator definitions can even be mixed together in Index arrays like the \inlinenohighlight{sites} array above, permitting easy setup of calculations for mixed systems such as spin chains of alternating $S=1/2$ and $S=1$ sites or models of alternating spin and boson sites.

\subsection{DMRG Observer System}

The DMRG code within ITensor is the most heavily used high-level feature of the library due to the continued popularity and staying power of the DMRG algorithm. Although ITensor's implementation of DMRG prints some useful details about the results of each sweep, such as the estimated energy (dominant eigenvalue) and typical bond dimension of the MPS being optimized, there are many situations where a user would like to customize the code further, such as to measure local observables throughout each sweep.

To make this customization process as easy as possible, the ITensor DMRG code accepts an optional \inlinenohighlight{observer} keyword argument which allows users to pass any object which is a sub-type of \inlinenohighlight{AbstractObserver}. This type should also have an overload of at least one of the methods \inlinenohighlight{measure!} and \inlinenohighlight{checkdone!} defined for it too. These methods can be defined in any way the user sees fit and have minimal requirements. Both are called by the ITensor DMRG code at each step of the DMRG algorithm.

The \inlinenohighlight{measure!} method gets passed a variety of properties describing the current state of the DMRG calculation, such as the number of the current sweep and location of the site(s) of the MPS whose local tensors are currently being optimized, and even the entire MPS itself. A customized \inlinenohighlight{measure!} function can use this information to produce a detailed snapshot of how the optimization is proceeding. One such use of the observer system in the past was to make animated movies of a DMRG calculation to be used in lectures.

The \inlinenohighlight{checkdone!} method can be defined if the user wants to set some criterion for the DMRG calculation to stop before all of the requested sweeps have been performed. Example criteria could include some measure of convergence, such as the energy variance, or an external signal from the user. 

\section{Applications of ITensor}

ITensor has been cited in approximately 450 research articles from 
2009 to 2021.\footnote{\mbox{List of papers citing ITensor}: \href{https://itensor.org/papers}{https://itensor.org/papers}} 
Below we highlight papers which show the diverse applications of ITensor. 
We expect to see ever wider applications in the future
as tensor network algorithms become more powerful for 
two- and three-dimensional systems, ab-initio Hamiltonians, and long-time dynamics \cite{White:2018q,Rakovszky:2020},
and as more applications of tensor networks
are developed in applied mathematics, computer science, and machine learning \cite{Oseledets:2010,Novikov:2015,pmlr-v108-rakhshan20a}.

\subsection{Equilibrium Quantum Systems}

The most common application area of tensor networks and the ITensor software to date
has been equilibrium quantum systems. A common starting point for understanding
equilibrium systems is through their ground state, and the DMRG algorithm
which launched the field of tensor networks is primarily a ground state finding method.
More recently, tensor network methods have been extended to study finite-temperature systems. 
Another important area of development in the field has been extending DMRG and MPS methods to handle ab initio systems such as in quantum chemistry, where details of continuum, atomic physics must be treated.

An excellent example of a ground-state study using ITensor is that of Keselman and Berg~\cite{Keselman:2015},
who used ITensor's DMRG algorithm to compute properties of a \emph{one-dimensional model of superconducting
electrons}. A detailed study of properties of finite-size systems, including of quantities at the
edge of open systems, supports the existence of a topological state of matter even in the absence
of a gap in the excitation spectrum.

The state-of-the-art efficiency of ITensor's DMRG codes makes it a powerful tool for studying
two-dimensional systems as well. DMRG remains one of the leading methods for studying two-dimensional
quantum systems even though it scales exponentially in the transverse system size.
In Refs.~\cite{Kallin:2014,Stoudenmire:2014}, Kallin, Gustainis, Johal, Stoudenmire, 
Melko, et al.\ used a combination of exact diagonalization, numerical linked cluster methods, 
and ITensor DMRG to 
obtain \emph{entanglement entropies associated with sharp corners} in the subsystem geometry for 
various quantum systems at their critical points. Based on the numerical results, a conjecture
was put forward for a universal scaling of this corner entanglement which was afterward supported by field theoretic methods \cite{Bueno:2015}.

An exemplary study using ITensor DMRG for a two-dimensional system of strongly-correlated
electrons is the work by Venderley and Kim \cite{Venderley:2019}, who  studied the hole-doped
\emph{Hubbard model on the triangular lattice}, finding a transition from p-wave to d-wave superconductivity
as the strength of on-site interactions increase.

ITensor has also been used for studying continuum electronic systems such as \emph{quantum chemistry} calculations
of hydrogen chains \cite{Stoudenmire:2017s,Motta:2017t,White:2019m};
for \emph{finite-temperature} studies, primarily in the context of the 
minimally entangled typical thermal state (METTS) 
algorithm \cite{White:2009,Stoudenmire:2010,Bruognolo:2017,Chen:2019};
and for calculations involving \emph{PEPS two-dimensional} tensor networks \cite{Jiang:2019,Hyatt:2019}.

\subsection{Dynamics of Quantum Systems}

Dynamical behavior of quantum systems or quantum systems out-of-equilibrium is currently an active
research area, where the flexibility and customizability offered by ITensor has been an excellent fit. 
Such customizability is important because there are many algorithms available for time-evolving 
quantum states \cite{Paeckel:2019}, most of which are
not totally black-box and require some care to use well. Frontier research problems also
involve a variety of settings, such as closed versus open systems, or evolution via Hamiltonians
versus circuits, as well as a wide range of measurements to be made of the state.

One paper typifying the use of ITensor for dynamics research, blending numerical results with theoretical
predictions, is that of Alba and Calabrese Ref.~\cite{Alba:2017}, who showed that for 
\emph{integrable systems}, such as the XXZ spin chain, one can accurately predict the entanglement entropy  
 at both short and long times. 

Nahum, Ruhman, Vijay, and Haah used ITensor 
in Ref.~\cite{Nahum:2017} to simulate
dynamics of quantum states evolved by \emph{random unitary circuits}, supporting their prediction that
the growth of entanglement entropy is governed by the KPZ universality class related to the classical statistical physics of surface growth.

Schreiber et al.\ used ITensor to simulate the \emph{dynamics
of cold atom experiments} in Ref.~\cite{Schreiber:2015}, 
obtaining good agreement with experimental observations of the difference
between the number of atoms in even versus odd minima of the external potential.

A rather different application of dynamical tensor network methods are as ``solver'' 
subroutines for the dynamical mean field theory (DMFT) algorithm, which can treat infinite-size
systems in two and three dimensions. A novel \emph{DMFT solver based on fork tensor 
network states} was proposed and demonstrated using ITensor by Bauernfeind, Zingl, et al.\ in Ref.~\cite{Bauernfeind:2017}, allowing DMFT methods to achieve greater resolution for electron
spectral functions and other benefits.

\subsection{Other Application Areas}

Historically tensor network methods have mainly been developed and applied within
condensed matter physics. But the recent decade has seen a major broadening in applications of tensor networks inside and outside of physics. These newer applications  
range from studying holographic dualities between physical theories \cite{Swingle:2012m,Pastawski_2015}
to computing high-dimensional integrals in applied mathematics \cite{Oseledets:2010,Dolgov:2020}.

An area where tensor network methods are becoming increasingly important is  
\emph{quantum computing}, where they can be used to perform efficient classical simulations
of quantum devices. Tensor networks offer important advantages 
such as linear scaling with the number of qubits.
The library \href{https://www.pastaq.org}{PastaQ} (available at \href{https://github.com/GTorlai/PastaQ.jl}{github.com/GTorlai/PastaQ.jl})
uses ITensor as a backend to offer tensor network methods not only for quantum simulation, but also optimization of quantum circuits, tomography of quantum systems and quantum processes, and more.


A rather different application area of tensor networks is applied mathematics and machine learning.
Here tensor decomposition methods have found many different uses, from compressing weight layers of neural networks \cite{Novikov:2015}, to recovering missing or corrupted data using partial information \cite{huang2019lowrank}. Machine learning is an area where ITensor has potential to be used much more 
in the future, and ITensor has already been used to \emph{investigate new models and algorithms} for machine learning,
including supervised \cite{NIPS2016_6211,reyes2020multiscale} and unsupervised \cite{Stoudenmire_2018} 
learning using models parameterized by tensor networks, and to investigate generalization of these models by studying 
synthetic data \cite{bradley2019modeling}.

\section{Benchmarks of ITensor Performance}

To ensure that ITensor offers state-of-the-art performance, 
we next present benchmark results of ITensor implementations of typical tensor network
algorithms and operations. 
One goal is comparing the performance of the C++ versus Julia implementations of ITensor, 
as Julia is a relatively new language whose
potential for high performance computing has not yet been fully verified in every domain. 
Other goals of the benchmarks include testing the scaling of algorithm implementations of ITensor and showing the relative benefits of multithreading. 
Finally, we discuss benchmarks of ITensors versus other tensor network libraries, which we make available
as an online resource, since all of the libraries involved are frequently updated and continually optimized.


All benchmarks shown here were carried out on a single workstation with four Intel Xeon Gold 6128 (Skylake) 3.4 GHz CPUs with six cores each. Times shown are ``wall'' or actual time, not CPU time. 
The BLAS and LAPACK distribution used for both the C++ and Julia calculations was Intel MKL. 
For the Julia ITensor benchmarks we used version 0.2.0 of ITensors.jl running on Julia version 1.6.1.
The benchmarks presented below are publicly available at: 
{\color{blue} \href{https://github.com/ITensor/ITensorBenchmarks.jl}{https://github.com/ITensor/ITensorBenchmarks.jl}}.

Before we present the benchmarks, here are the high-level conclusions we draw from them:
\begin{itemize}
    \item At least for the domain of tensor network algorithms, Julia is very competitive with C++ as a high-performance programming language.
    \item Some ITensor algorithms, especially those involving block sparse tensors, are currently fastest in the Julia implementation due to recent optimization efforts made there. Though most of these optimizations can be carried out in C++ too, the productivity of the Julia language and its superior libraries and tooling makes optimizations easier to identify and implement.
\end{itemize}

We again emphasize that the Julia version of ITensor is written entirely in the Julia language, without needing to
perform any low-level operations in systems languages such as C++ as is often necessary in languages like Python to achieve high performance. Of course certain external libraries we use, such as BLAS and LAPACK, are written in other languages such as Fortran, but such libraries are standard and widely used by many tensor libraries including both the C++ and Julia implementations of ITensor.

\subsection{Comparison of Julia and C++ Implementations of ITensor}

First we present a set of benchmarks comparing the performance of the C++ and Julia implementations of ITensor on seminal tensor network algorithms.

As a first comparison between the C++ and Julia implementations of ITensor,
a simple but powerful tensor network algorithm is the tensor renormalization group (TRG) \cite{Levin:2007,Gu:2008}, which computes properties of classical statistical mechanics models at finite temperature through a decimation procedure. Each step of TRG essentially consists of contracting four tensors together into a single tensor, then performing a truncated factorization of that tensor. Below we present benchmarks of the TRG algorithm in the C++ and Julia version of ITensor, using dense tensors only, and showing calculations with 1, 4, and 8 threads used by the BLAS library within the tensor contraction steps and for different maximum bond dimensions used during the truncation steps:

\begin{center}
\includegraphics[width=0.8\columnwidth]{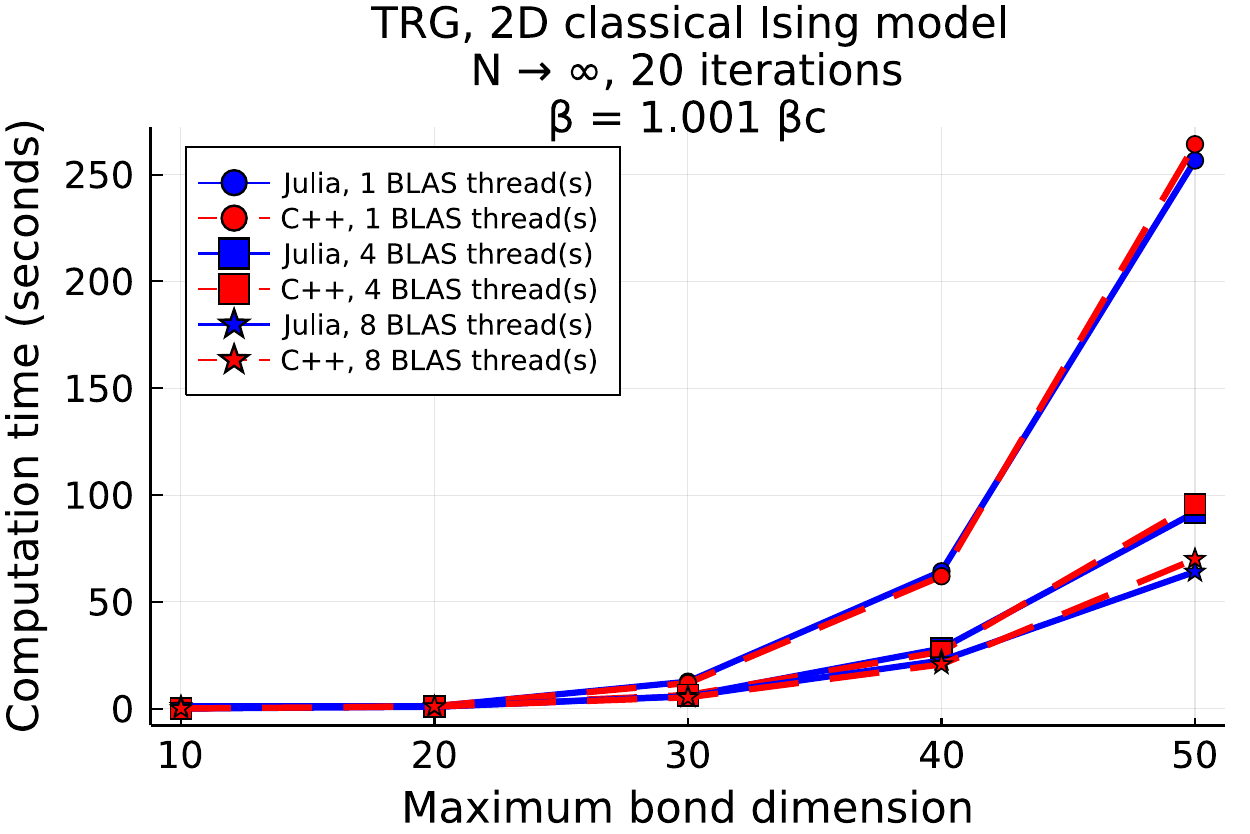}
\end{center}

From the results, we can see that the C++ and Julia implementations have very similar performance, with the C++ version performing slightly better at bond dimension 40 and the Julia version performing better at bond dimension 50. The BLAS and LAPACK threading is clearly effective for speeding up these contraction-dominated calculations.

Another algorithm used to study classical statistical models, as well as to contract infinite PEPS tensor networks, is the corner transfer matrix renormalization group (CTMRG) \cite{Nishino:1996,Nishino:1997,Fishman:2018}. 
The CTMRG algorithm decimates a contracted network of tensors by absorbing bulk tensors into boundary tensors and computing new boundary tensors at each step. Below we show the benchmark results for CTMRG using dense tensors:

\begin{center}
\includegraphics[width=0.8\columnwidth]{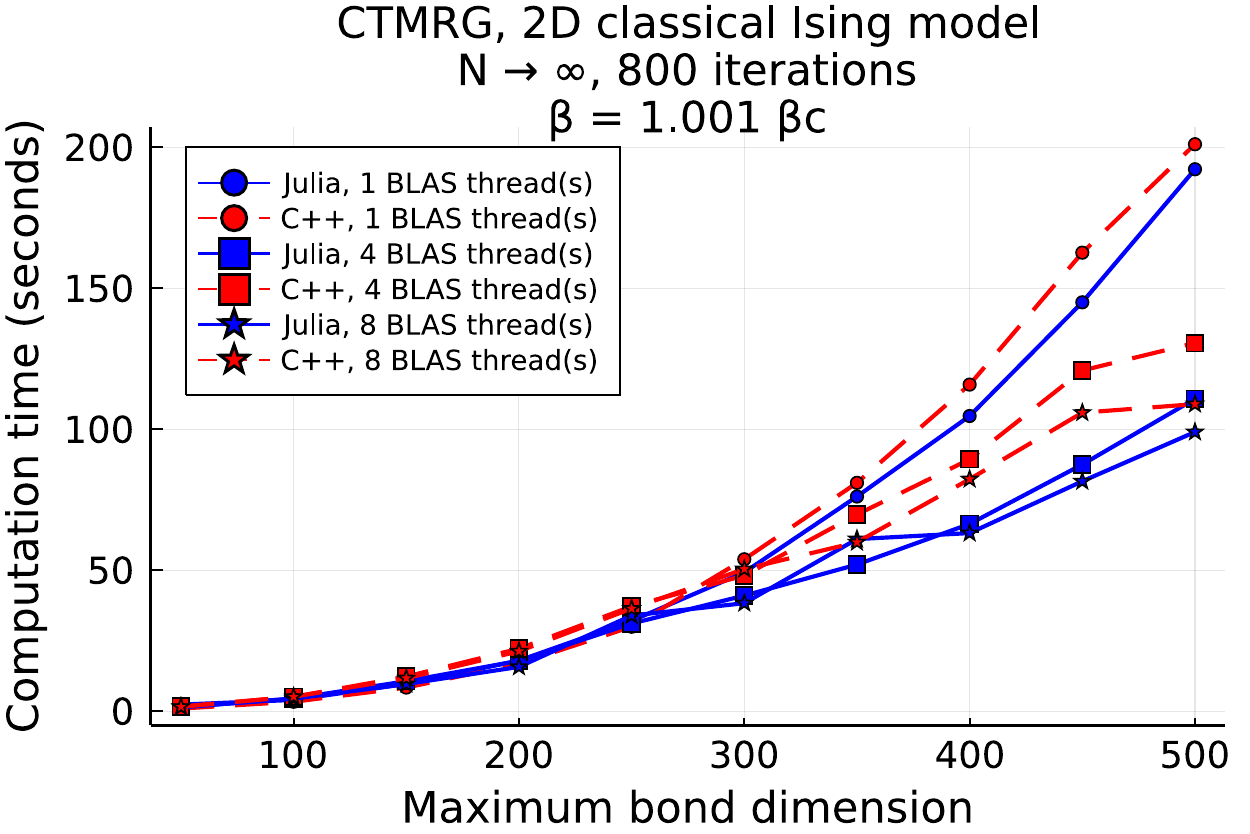}
\end{center}

Here the Julia implementation is consistently faster for a wide range of larger bond dimensions of the boundary tensors. Allowing the BLAS to use four threads gives a speedup, but using eight threads gives little additional speedup. The relative performance as a function of BLAS threads is similar between the C++ and Julia codes, showing how the effectiveness of BLAS multithreading is dependent on the system studied and algorithm used. Speedups of the Julia versus the C++ calculations are likely due to improved dense tensor permutation libraries, specifically \href{https://github.com/Jutho/Strided.jl}{Strided.jl}, used in the Julia version.

Now we turn to benchmarks of the density matrix renormalization group (DMRG) algorithm. DMRG calculations are the most common application of the ITensor library. We will also use DMRG as a setting to study the effect of conserving quantum numbers, resulting in block sparse tensors.

First we benchmark the simplest application of DMRG: a one-dimensional spin chain, with no quantum number conservation, that is, dense tensors:

\begin{center}
\includegraphics[width=0.8\columnwidth]{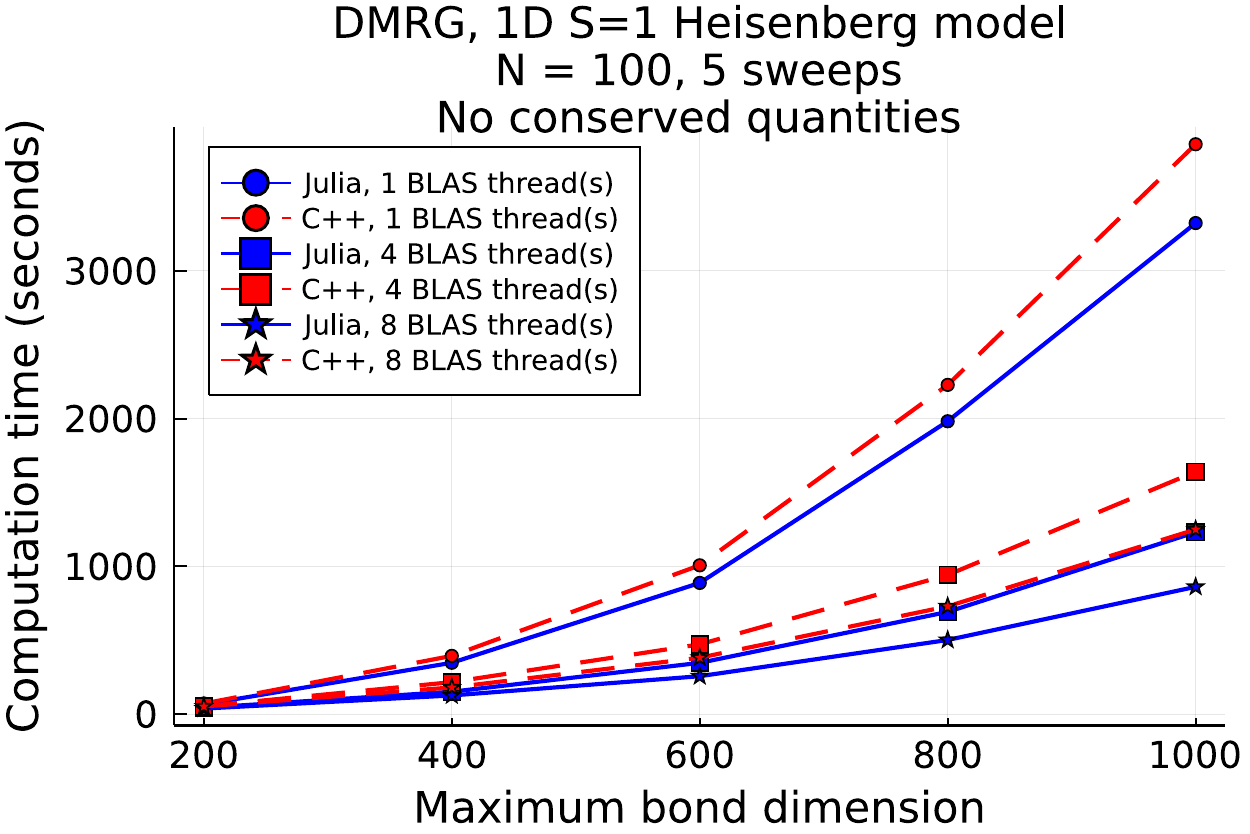}
\end{center}

The relatively better performance of the Julia version over the C++ implementation is similar to that for CTMRG, which is sensible as the details of both algorithms are similar.

Next we consider DMRG for the same system, but conservation of the total $S^z$ spin quantum numbers and taking advantage of the resulting tensor block sparsity:

\begin{center}
\includegraphics[width=0.8\columnwidth]{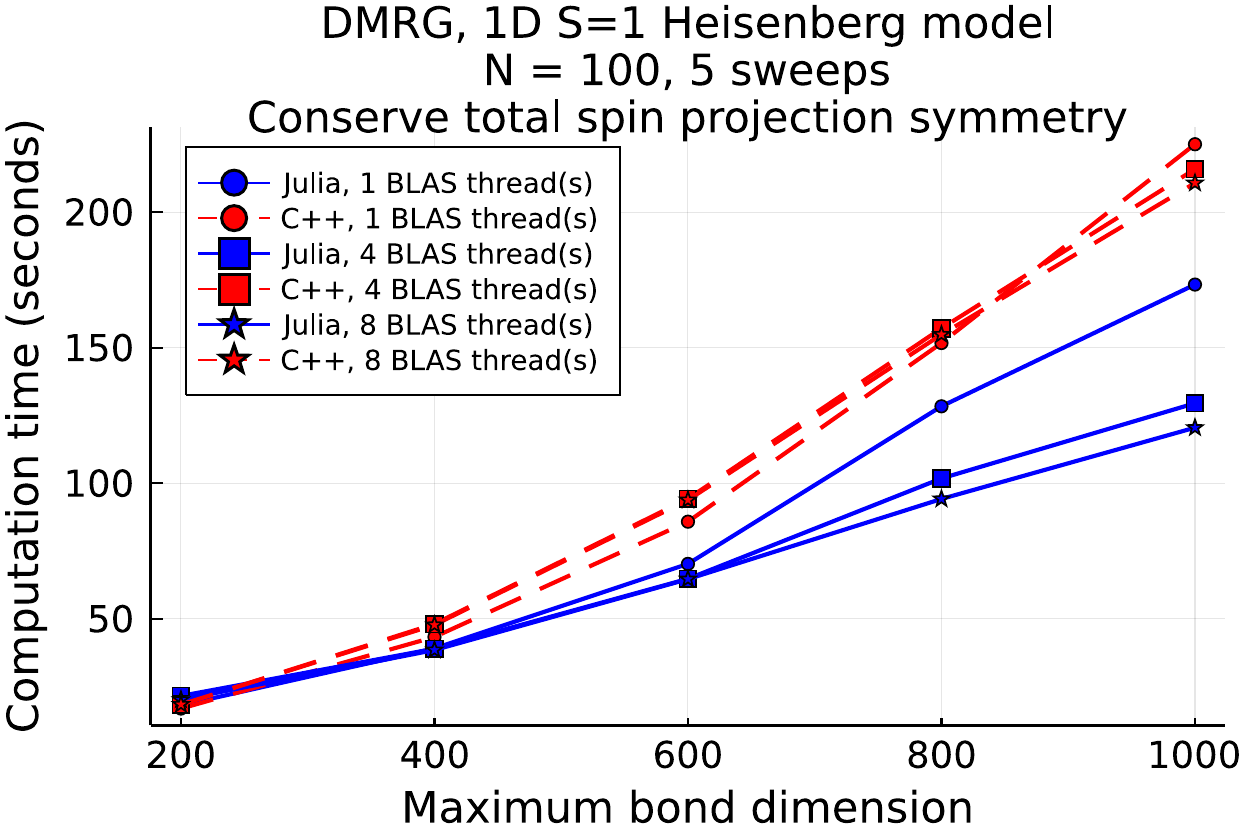}
\end{center}

The results above show that the handling of block sparse tensors is currently much more efficient in the Julia version of ITensor versus the C++ version. This is the result of an extensive recent optimization effort, using techniques such as storing the locations of the non-zero blocks in a dictionary data structure instead of an array and optimizing contractions of small blocks. An interesting contrast of block sparse calculations versus dense calculations is that BLAS multithreading is much less effective in the block sparse case, which is likely because many of the blocks are much smaller than the overall tensor dimension, leading to smaller matrices being multiplied at the BLAS level.

Finally we benchmark the DMRG algorithm for a quasi-two-dimensional system treated by wrapping an MPS on a cylinder. Here we use the example of the Hubbard model with $U/t=8$ and conservation of both the total $S^z$ and particle number symmetries:

\begin{center}
\includegraphics[width=0.8\columnwidth]{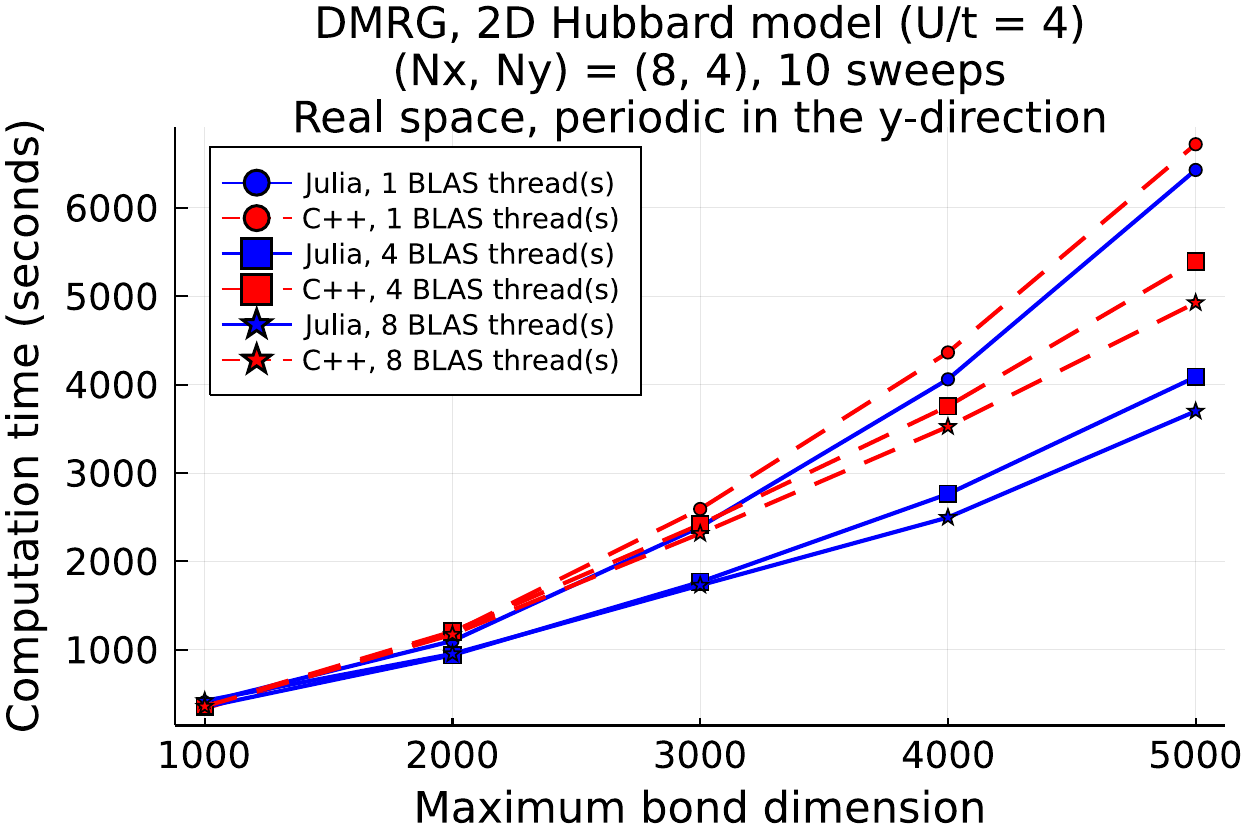}
\end{center}

While the Julia version also outperforms the C++ version for this system, the single-threaded case is similar for both code versions, perhaps due to certain larger non-zero tensor blocks. 

A technique to sparsify the tensors more in the context of two-dimensional DMRG calculations is to also conserve the momentum quantum number $k_y$ in the y-direction, or periodic direction around the cylinder \cite{Motruk:2016}. By using that technique in the following benchmarks of the same two-dimensional Hubbard system, we can see that the overall time needed is reduced and the better-optimized block sparse operations in the Julia version give an even larger advantage:

\begin{center}
\includegraphics[width=0.8\columnwidth]{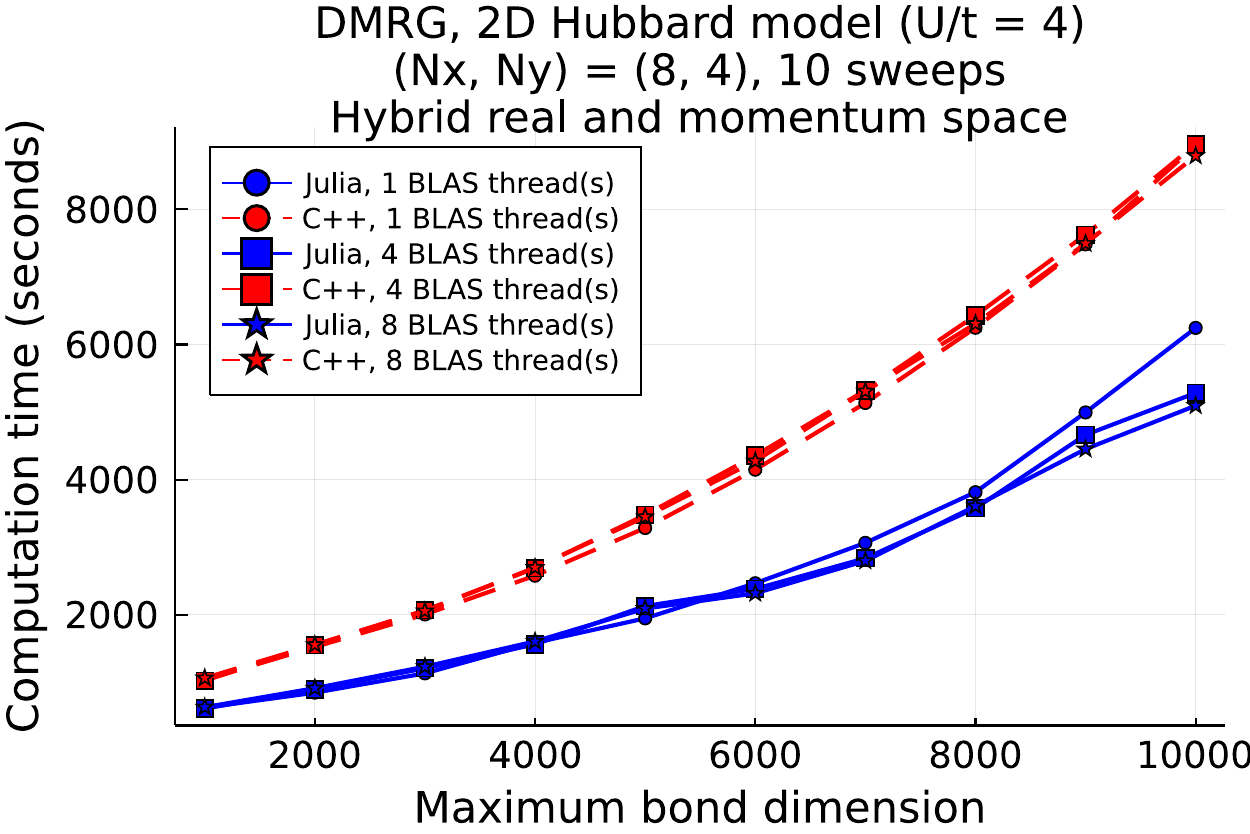}
\end{center}

Finally, the block-sparse structure of quantum-number conserving tensors gives an opportunity for performing contractions of the non-zero blocks in parallel. We offer multithreading over block-sparse tensor contractions in both the C++ and Julia versions of ITensor. Turning on this feature and using different numbers of threads for 2D DMRG calculations gives the following timings:
\begin{center}
\includegraphics[width=0.8\columnwidth]{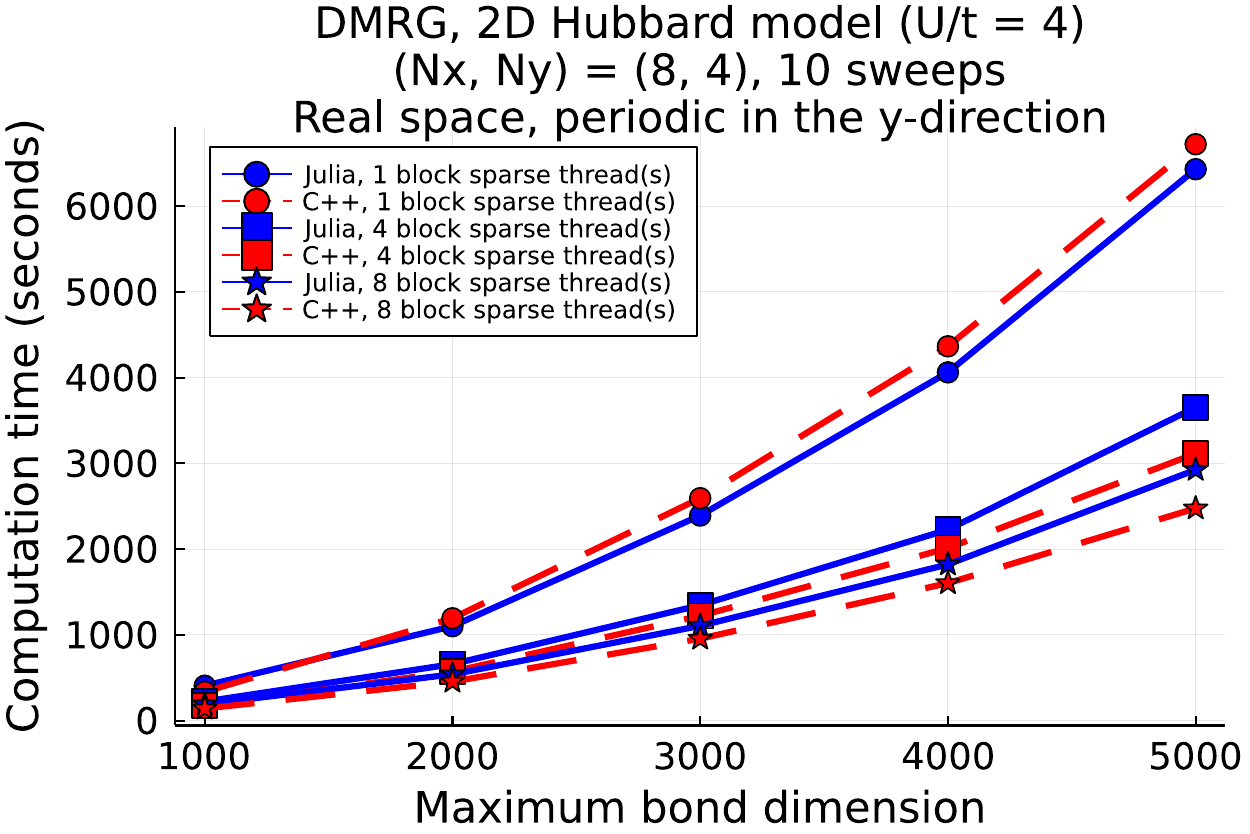}
\end{center}
where we see a speedup of between 1.5x to 2x compared to case using no block-sparse threading.
Though the single-threaded Julia implementation is slightly more efficient, the
multithreading is more effective in the C++ implementation, possibly because the native Julia multithreading has a higher overhead than the OpenMP multithreading we use in C++. We plan to investigate the discrepancy in more detail.

To conclude this section, we note that the C++ implementation of ITensor, including both its tensor contraction routines and implementations of algorithm such as DMRG are already highly optimized, nearing state-of-the-art performance. So the even better performance of the Julia version of ITensor is a non-trivial outcome. The Julia version was originally modeled on the  C++ implementation, but recent optimization efforts supported by Julia's more productive programming environment currently put it well ahead.

\subsection{Benchmarks of ITensor Versus Other Software}

It is important to determine how the performance of ITensor compares to other leading software.
For this purpose, we have performed benchmarks comparing the Julia version of ITensor to the \href{https://tenpy.readthedocs.io/en/latest/}{TeNPy} high-performance tensor network library, which is implemented in a combination of Python and C++  with a Python interface \cite{tenpy}.

However, because both ITensor and TeNPy are continually being optimized and developed, and due to subtleties of comparing different implementations of algorithms such as DMRG, we have opted not to present a definitive set of benchmarks here, but rather to host these on an external site where the results and underlying codes can be periodically updated. The latest TeNPy and ITensor benchmarks can be viewed at the following link: {\color{blue} \href{https://itensor.github.io/ITensorBenchmarks.jl/dev/tenpy_itensor/index.html}{ITensorBenchmarks TeNPy and ITensor Comparisons}}. 

To summarize the results of this ongoing benchmark effort, we found first of all a number of implementation differences that can inform the design and default choices of each library. For example, TeNPy by default uses a sparse representation of the Hamiltonian which we found typically speeds up DMRG significantly, so we have now implemented a similar capability in ITensor through a function called \inlinenohighlight{splitblocks}, though whether using leads to better performance depends on the system, so we have not currently made it the default.
Another difference is that TeNPy's DMRG implementation (as of version 0.8.4) performs more Lanczos steps within each step of DMRG compared to ITensor, which generally results in longer running times for a fixed number of DMRG sweeps. But this number is only a default setting and can be adjusted by the user. Once the algorithmic details and external dependencies (such as the BLAS library used) were made as similar as possible, we found both libraries gave comparable performance.

In the future, we plan to not only expand the set of algorithms used in the benchmark, but also to set up an automatic benchmarking system, and to include other software in the comparisons.

\section{Future Directions}

Although the ITensor library already offers high performance and
powerful features for implementing any tensor network algorithm, many
improvements and optimizations are planned or already under way.
Here we discuss the main features under development, though some may
take a different form when implemented.

A high-priority feature is support for automatic differentiation (AD).
This technique has been popularized for applications in machine learning
and neural networks, but has recently been demonstrated to work well for
tensor networks too. For example, AD can be used for state-of-the-art
infinite PEPS calculations and for calculating critical properties of
classical systems \cite{Liao:2019}.
In addition, it has proven useful for optimizing tensor networks with
unitary/isometric constraints like quantum circuits, MERA, and
gauged MPS \cite{Evenbly:2014,Luchnikov:2021,Hauru:2021,Geng:2021}
as well as for computing excitations and structure factors of MPS and
PEPS \cite{Wei-Lin:2021,Ponsioen:2021}.
The unique index system and generic high level interface makes ITensor
ideal for defining differentiation through a variety of ITensor
operations.
Julia's ChainRules.jl\cite{White:ChainRules:2021,White:ChainRulesCore:2021} package
can be used to define basic reverse and forward
mode differentiation rules independent of the particular AD framework.
In conjunction with source-to-source AD frameworks available in Julia such as
Zygote.jl \cite{Innes:2018} which has high coverage for differentiating through
most native Julia language features, a basic ITensor AD system involving differentiating
through a surprising number of ITensor operations can be written in only
a few lines of code.
Our use of ChainRules will allow us to target next generation AD systems being
developed in Julia such as 
\href{https://github.com/JuliaDiff/Diffractor.jl}{Diffractor.jl}.
Using this system, we have prototypes for using AD to optimize a variety of
tensor network applications, such as gradient optimization of MPS, variational
circuit optimization, and PEPS.
We plan to extend our set of rules and coverage of ITensor operations (for example
better support for differentiating tensor factorizations and MPS/MPO operations),
incorporate high level support for using AD to optimize ITensor networks
with unitary constraints, etc.
In addition, we are investigating adding features for computing higher order
derivatives of tensor networks using backends like AutoHOOT \cite{Ma:2020}.

Another high-priority feature is automatic support for fermionic
Hilbert spaces. Systems of fermions are foundational for physics
applications of tensor networks, and are the most common type of
system studied in condensed matter physics. Currently, the only
automatic support for fermions in ITensor is within the OpSum/AutoMPO
system, which relies on lookup tables of operator names designated
as anti-commuting. That approach works well for many matrix product
state calculations, but leads to a confusing experience for users
when some parts of the library handles fermions automatically yet
other parts of the calculation require manually introducing
Jordan-Wigner string operators, such as when computing certain
correlation functions or when  using
higher-dimensional networks such as PEPS. We are therefore experimenting
with a system that introduces fermionic properties at the level of
tensor indices, where index permutations result in a minus sign
if odd-parity QN subspaces undergo an odd-parity permutation.
Our ambitious goal is for calculations involving
fermions to work with exactly the same code as for bosonic degrees of
freedom. Even if some manual steps are occasionally required, this new
fermion system could still be very useful.

Following the completion of the fermion system, support for other 
types of symmetries and non-trivial vector spaces is an important future direction.
In particular, support of non-Abelian symmetries such as $SU(2)$ 
will be a very powerful feature for variants of the Heisenberg and Hubbard models 
and for electronic structure Hamiltonians such as in quantum chemistry applications.

More sophisticated optimizations of tensor contraction sequences is
another future direction for ITensor.
We currently have a backend for optimizing the contraction sequence
of ITensors, for example to determine that the optimal sequence of 
a contraction like \inline{A*B*C*D} is \inline{(A*(B*C))*D},
based on the algorithm introduced in Ref.~\cite{Pfeifer:2014}.
This can be enabled for every contraction with a global flag or
for a specific contraction with a keyword argument, and additionally
a custom sequence can be provided of the form \inline{[[1,[2,3]],4]}.
We are also developing tools for visualizing tensor networks which are enabled by annotating a tensor contraction with a macro, for example \inline{@visualize A*B*C*D}. We plan to provide a variety of backends, such as a text output and an interactive output based on Makie.jl\cite{Makie}.

We are also developing tools for visualizing tensor networks which are enabled by annotating a tensor contraction with a macro, for example \inline{@visualize A*B*C*D}. We plan to provide a variety of backends, such as a text output and an interactive output based on Makie.jl\cite{Makie}.
This will make it easier to visualize a contraction sequence and debug code.
We would like to provide alternative contraction sequence optimization backends
like \href{https://github.com/jcmgray/cotengra}{CoTenGra} 
\cite{Gray:2021} which could be used to find contraction sequences
for larger tensor networks than our current implementation.
In addition, we are investigating incorporating general approximate
contraction algorithms like those introduces in Refs.~\cite{Pan:2020,Chubb:2021}.

We soon plan to offer first-class support for infinite
MPS and MPO algorithms, with preliminary work nearly completed in the
currently separate package 
\href{https://github.com/ITensor/ITensorInfiniteMPS.jl}{ITensorInfiniteMPS.jl}.
This will include the latest developments
in obtaining dominant and sub-dominant eigenvalues and MPS eigenvectors
of infinite MPOs, using algorithms such VUMPS \cite{Zauner-Stauber:2018}
and MPS tangent-space methods \cite{TangentSpaceMethods}, 
as well as obtaining canonical forms of infinite MPS and MPOs and
applying infinite MPOs to infinite MPS \cite{Vanhecke:2021}. This will
all be offered with the same level of convenience as the 
currently available finite MPS and MPO methods, including an infinite
version of OpSum/AutoMPO.

We plan to continue developing GPU support throughout the library.
Currently, only dense tensor operations can be performed on GPU,
so an initial goal will be to support block sparse tensor operations on GPU.
More broadly, we plan to make GPU support a first-class feature,
with the eventual goal that most code written for ITensors on CPU
can work directly for ITensors on GPU with high performance and minimal
user effort, including code that uses automatic differentiation.

Last but not least, we hope to offer more high-level features for PEPS
(two-dimensional tensor network) calculations. Algorithms and methods
for optimizing PEPS have reached a point of maturity such that there are
now a handful of essentially standard approaches, such as variational
iPEPS \cite{Vanderstraeten:2016,corboz2016variational,Liao:2019} and fixed-point
methods for computing PEPS environment tensors \cite{Fishman:2018}. Many of these
algorithms will be provided with ITensor in the future, and in particular
leverage tools we are developing for a general tensor network interface,
automatic differentiation, and contraction sequence optimization.

\section*{Acknowledgements}

We thank Johannes Hauschild for many discussions about the TeNPy software and for taking significant time to work with us to provide and develop benchmark codes. We thank Nils Wentzell for providing expertise and help regarding a custom Python environment on the Flatiron Institute computing cluster, as well as help designing the multithreading strategy for threaded block sparse contractions.

Key contributors to ITensor include: Katharine Hyatt for developing a GPU-accelerated backend for the ITensors.jl package;
\footnote{\mbox{ITensorGPU:} \href{https://github.com/ITensor/ITensors.jl/tree/main/ITensorGPU}{https://github.com/ITensor/ITensors.jl/tree/main/ITensorGPU}} Anna Keselman for contributing a major improvement to the OpSum/AutoMPO system which handles long-range interactions and multi-site operators; Thomas E. Baker for expanding and improving the ITensor documentation, in particular the tutorials. Thanks to Jing Chen, Ying-Jer Kao, John Terilla, and Tyler Bryson for discussions about automatically handling fermion signs. We also thank Jing-Guo Liu for helping us to generalize the tensor contraction backend of ITensors.jl to handle more arbitrary number types.

Significant contributions and bug fixes to the C++ version of ITensor were made by 
Anna Keselman, Mingru Yang, Jack Kemp, Kyungmin Lee, Tatsuto Yamamoto, Juraj Hasik, Benedikt Bruognolo, 
Jose Lado, Hoi Hui, Lars-Hendrik Frahm, Lucas Vieira, Markus Wallerberger, Miles Chen, 
Yevgeny Bar-Lev, Jessica Alfonsi, Chuang Xi, and Andrey Antipov. We would also like to thank Nils Wentzell, Alex Wietek, and Daniel Bauernfeind for their help designing and testing block sparse multi-threading with OpenMP.

Significant features and bug fixes to the initial release of ITensors.jl (the Julia version of ITensor) were contributed by Katharine Hyatt, Ori Alberton, Christopher White, Jan Schneider, Alvaro Rubio-Garcia, Yiqing Zhou,  Michael Abbott,  Nicolau Werneck, Michael Sven Ferguson, Nick Robinson, and Amartya Bose.

\paragraph{Funding information}
SRW acknowledges the support of the U.S.\ Department of Energy under grant DE-SC0008696. 
ITensor was initiated through the generous support of the DOE under award DE-SC0008696 and the NSF under award DMR-1812558, both of which continue to support the efforts of Steven R. White and his group.
We are grateful for ongoing support through the Flatiron Institute, a division of the Simons Foundation.

\begin{appendix}

\section{Full Code Examples}

In addition to the code examples below, we include an extensive and growing set of examples 
as part of our source code distribution at the following link: {\color{blue} \href{https://github.com/ITensor/ITensors.jl/tree/main/examples}{ITensor Code Examples}}.

\subsection{Contraction Example \label{sec:julia_example_1}}

To show a fully working example of contracting two ITensors with a complicated index structure,
consider the following code\footnote{Collections of indices can be made with a more compact syntax \inline{a,b,c,d,i,j = Index.((3,2,4,5,2,6),("a","b","c","d","i","j"))}, which makes use of Julia's built in broadcast (\inlinenohighlight{.}) syntax.}:

\begin{jlcodeblock}[frame=single,float=!h,label=julia_contraction]
using ITensors

function main()
  a = Index(3,"a")
  b = Index(2,"b")
  c = Index(4,"c")
  d = Index(5,"d")
  i = Index(2,"i")
  j = Index(6,"j")

  A = randomITensor(a,b,d,c)
  B = randomITensor(i,d,j)
  
  C = A * B
  
  @show hasinds(C,a,b,c,i,j)

  return C
end

main()
\end{jlcodeblock}
The contraction computed by this code can be expressed by the following diagram:
\begin{center}
\includegraphics[width=0.6\columnwidth]{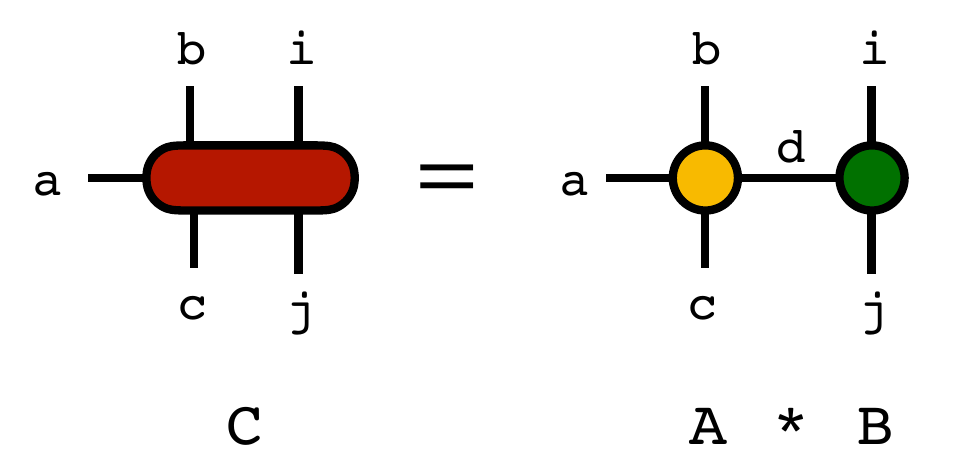}
\end{center}
Note that the \inlinenohighlight{Index} tags such as \inlinenohighlight{"a"},\inlinenohighlight{"b"},\inlinenohighlight{"c"}, etc. are not required for 
this code to function properly, but in this context are just for making the indices
easier to identify when printed.

The line of code 
\begin{jlcodeblock}
@show hasinds(C,a,b,c,i,j)
\end{jlcodeblock} 
shows the output of the \inlinenohighlight{hasinds} function which checks that the ITensor \inlinenohighlight{C}
has all of the indices \inlinenohighlight{a,b,c,i,j}. The code above will output
\begin{jlcodeblock}
hasinds(C,a,b,c,i,j) = true
\end{jlcodeblock}

\subsection{DMRG Example \label{sec:julia_example_2}}

The following code example shows the use of higher-level features of the ITensor Library
to compute the ground-state wavefunction of the $S=1/2$ Heisenberg
quantum spin chain model using the density matrix renormalization group (DMRG) 
algorithm:

\begin{jlcodeblock}[frame=single,float=!h,label=julia_dmrg]
using ITensors

function main(N)
  sites = siteinds("S=1/2",N)

  os = OpSum()
  for j=1:N-1
    os += "Sz",j,"Sz",j+1
    os += 1/2,"S+",j,"S-",j+1
    os += 1/2,"S-",j,"S+",j+1
  end
  H = MPO(os,sites)

  psi0 = randomMPS(sites; linkdims=10)

  sweeps = Sweeps(5)
  setmaxdim!(sweeps, 10,20,100,100,200)
  setcutoff!(sweeps, 1E-11)

  energy, psi = dmrg(H,psi0, sweeps)
  println("G.S. energy = $energy")
  return energy, psi
end

energy, psi = main(100)
\end{jlcodeblock}
\newpage 
A typical output of this code is:
\begin{center}
\includegraphics[width=0.8\columnwidth]{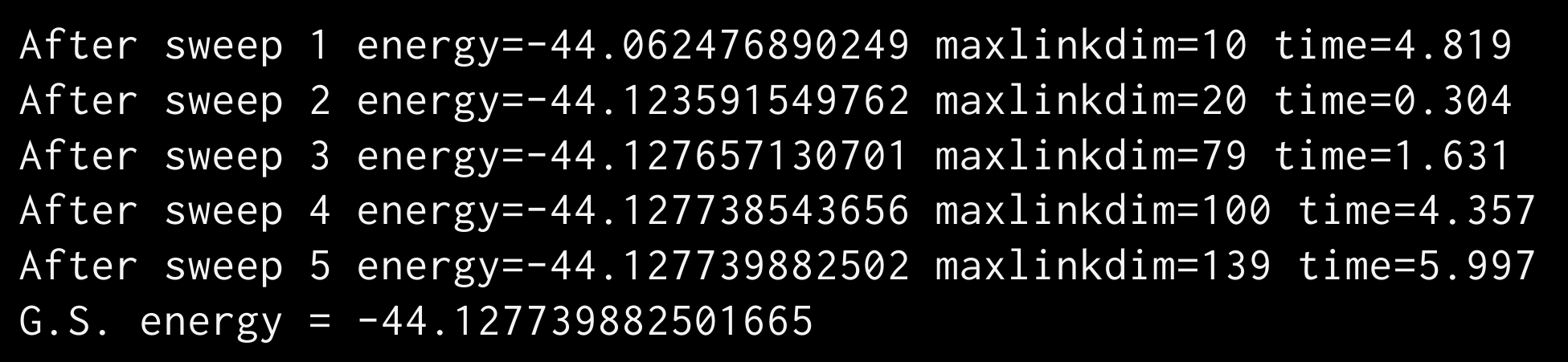}
\end{center}
where note that the longer time in the first sweep includes compilation time.
Brief explanations of the major steps of the above code are:
\begin{itemize}
\item Construct an array of $N=100$ \inlinenohighlight{Index} objects corresponding to $S=1/2$ spins (which are dimension-2 \inlinenohighlight{Index} objects labeled by the tag \inlinenohighlight{"S=1/2"}).
\item Input the terms of the one-dimensional Heisenberg Hamiltonian into an \inlinenohighlight{OpSum} object.
\item Construct an MPO \inlinenohighlight{H} out of the \inlinenohighlight{OpSum}.
\item Construct a random MPS \inlinenohighlight{psi0} of bond dimension 10.
\item Create a \inlinenohighlight{Sweeps} struct which indicates that five sweeps of the DMRG algorithm are to be performed, with various maximum bond dimensions allowed for each sweep and a truncation error cutoff of $10^{-11}$ throughout.
\item Run the DMRG algorithm, which returns the ground-state energy and ground-state wavefunction MPS.
\end{itemize}


\section{ITensor Implementation and Interface in the C++ Language}

In this appendix, we give code examples for the C++ version of ITensor to show the similarities to and differences
from the Julia version.

\subsection{C++ Contraction Example}

Here we show the same example of contracting two ITensors with a complicated index structure as
in the previous Appendix section \ref{sec:julia_example_1}.
Consider the following code:

\begin{jlcodeblock}[frame=single,float=!h,label=cpp_contraction]
#include "itensor/all.h"
#include "itensor/util/print_macro.h"
using namespace itensor;

int main()
  { 
  auto a = Index(3,"a");
  auto b = Index(2,"b");
  auto c = Index(4,"c");
  auto d = Index(5,"d");
  auto i = Index(2,"i");
  auto j = Index(6,"j");

  auto A = randomITensor(a,b,c,d);
  auto B = randomITensor(i,d,j);

  auto C = A * B;
  
  Print(hasInds(C,a,b,c,i,j));
  }   
\end{jlcodeblock}

By comparing to the Julia language example \ref{sec:julia_example_1}, one can see that the C++ code
above is very similar with the main differences being the use of \inlinenohighlight{include} statements to import the
library headers, the use of the C++ keyword \inlinenohighlight{auto} on lines 
of code that result in the definition of a new variable, and semicolons terminating each line of 
procedural code. 
The last line uses a macro \inlinenohighlight{Print}
provided by ITensor, which has a similar behavior to the Julia \inlinenohighlight{@show} macro and 
which in this case generates the output:

\begin{jlcodeblock}
Print(hasInds(C,a,b,c,i,j)) = true
\end{jlcodeblock}

\subsection{C++ DMRG Example}

Here we show the same example of a DMRG calculation as
in the previous Appendix section \ref{sec:julia_example_2}.
Consider the following code:

\begin{jlcodeblock}[frame=single,float=!h,label=cpp_dmrg]
#include "itensor/all.h"
using namespace itensor;

int main()
  {
  auto N = 100;

  auto sites = SpinHalf(N,{"ConserveQNs=",false});

  auto ampo = AutoMPO(sites);
  for(auto j : range1(N-1))
    { 
    ampo += "Sz",j,"Sz",j+1;
    ampo += 0.5,"S+",j,"S-",j+1;
    ampo += 0.5,"S-",j,"S+",j+1;
    } 
  auto H = toMPO(ampo);

  auto psi0 = randomMPS(sites,10);

  auto sweeps = Sweeps(5);
  sweeps.maxdim() = 10,20,100,100,200;
  sweeps.cutoff() = 1E-11;

  auto [energy, psi] = dmrg(H,psi0,sweeps,{"Quiet=",true});
    
  println("G.S. energy = ",energy);
  }
\end{jlcodeblock}

\newpage

By comparing to the Julia language example \ref{sec:julia_example_2}, one can see that the codes
are again rather similar overall. Some key differences beyond the ones mentioned for the contraction example
include that the site \inlinenohighlight{Index} arrays (``site sets'') in the C++ version include quantum number
information by default, which we turn off in this example, and the \inlinenohighlight{dmrg} routine outputs 
much more information by default, so we pass the named argument \inline{\{"Quiet=",true\}}. These
two parts of the code highlight a custom named-argument system developed for the C++ version of ITensor
which could be more generally useful in other C++ codes and which we plan to release as a separate
library in the future.

\end{appendix}




\bibliography{main.bib}

\nolinenumbers

\end{document}